\documentclass[journal=jacsat,manuscript=article, layout=twocolumn]{achemso}

\setkeys{acs}{etalmode=truncate}
\setkeys{acs}{maxauthors=10}
\usepackage{chemformula} 
\usepackage[T1]{fontenc} 
\usepackage{mhchem}
\usepackage{chemfig,amsmath,amssymb,amsfonts}%
\usepackage{braket}

\newcommand{\overlap}[2]{\langle #1 | #2 \rangle}

\usepackage{graphicx}
\usepackage{xr}

\SectionNumbersOn

\author{Henry \textcolor{black}{J.} Thompson}
\affiliation[Soton]
{School of Chemistry and Chemical Engineering, University of Southampton, Southampton, SO171BJ, United Kingdom}
\altaffiliation{These authors contributed equally to this work.}
\author{Matteo Bonanomi}
\affiliation{Dipartimento di Fisica, Politecnico di Milano, 20133, Milano, Italy}
\alsoaffiliation{CNR - Istituto di Fotonica e Nanotecnologie (IFN), 20133, Milano, Italy}
\altaffiliation{These authors contributed equally to this work.}
\author{Jacob Pedersen}
\affiliation{Department of Chemistry, Technical University of Denmark, Kgs. Lyngby, DK-2800, Denmark}
\alsoaffiliation{Department of Chemistry, Norwegian University of Science and
Technology, Trondheim, N-7491, Norway}
\author{Oksana Plekan}
\affiliation[Elettra]
{Elettra - Sincrotrone Trieste S.C.p.A., Basovizza,
34149, Trieste, Italy}
\author{Nitish Pal}
\affiliation[Elettra]
{Elettra - Sincrotrone Trieste S.C.p.A., Basovizza,
34149, Trieste, Italy}
\author{Cesare Grazioli}
\affiliation[CNRIOM]
{CNR - Istituto Officina dei Materiali (IOM), Basovizza, 34149, Trieste, Italy}
\author{Kevin C. Prince}
\affiliation[Elettra]
{Faculty of Mathematics and Physics, Department of Surface and
Plasma Science, Charles University, Prague, 18000,
Czech Republic}
\alsoaffiliation{Elettra - Sincrotrone Trieste S.C.p.A., Basovizza,
34149, Trieste, Italy}
\author{Bruno N. C. Tenorio}
\affiliation{Department of Chemistry, Technical University of Denmark, Kgs. Lyngby, DK-2800, Denmark}
\author{Michele Devetta}
\author{Davide Faccial\`a}
\author{Caterina Vozzi}
\affiliation[CNRIFN]
{CNR - Istituto di Fotonica e Nanotecnologie (IFN), 20133, Milano, Italy}
\author{Paolo Piseri}
\affiliation{Dipartimento di Fisica “Aldo Pontremoli”, Universit\'a degli Studi di
Milano, 20133, Milano, Italy}
\author{Miltcho B. Danailov}
\author{Alexander Demidovich}
\author{Alexander D. Brynes}
\author{Alberto Simoncig}
\affiliation[Elettra]
{Elettra - Sincrotrone Trieste S.C.p.A., Basovizza,
34149, Trieste, Italy}
\author{Marco Zangrando}
\affiliation[Elettra]
{Elettra - Sincrotrone Trieste S.C.p.A., Basovizza,
34149, Trieste, Italy}
\alsoaffiliation
{CNR - Istituto Officina dei Materiali (IOM), Basovizza, 34149, Trieste, Italy}
\author{Marcello Coreno}
\affiliation[CNRISM]
{CNR - Istituto di Struttura della Materia (ISM), Basovizza, 34149, Trieste, Italy}
\author{Raimund Feifel}
\author{Richard J. Squibb}
\affiliation[Gothe]
{Department of Physics, University of Gothenburg, Gothenburg, 41296, Sweden}
\author{David M. P. Holland}
\affiliation[Dare]
{Science and Technology Facilities Council (STFC), Daresbury Laboratory, Warrington, WA4 4AD, UK}
\author{Felix Allum}
\affiliation[SLAC]
{Linac Coherent Light Source, SLAC National Accelerator Laboratory, Menlo Park, 94025, California, USA}
\author{Daniel Rolles}
\affiliation[KSU]
{J.R. Macdonald Laboratory, Department of Physics, Kansas State University, Manhattan, 66506, Kansas, USA}
\author{Piero Decleva}
\affiliation{Dipartimento di Science Chimiche e Farmaceutiche, Universit\'a degli
Studi di Trieste, Trieste, 34127, Italy}
\author{Michael S. Schuurman}
\affiliation{National Research Council Canada, Ottawa, K1A0R6, Ontario, Canada}
\alsoaffiliation{Department of Chemistry and Biomolecular Sciences, University of
Ottawa, Ottawa, K1N6N5, Ontario, Canada}
\author{\textcolor{black}{Ruaridh Forbes}}
\affiliation[SLAC]
{Linac Coherent Light Source, SLAC National Accelerator Laboratory, Menlo Park, 94025, California, USA}
\alsoaffiliation{Department of Chemistry, University of California, Davis, 95616, California, USA}
\author{Sonia Coriani}
\affiliation{Department of Chemistry, Technical University of Denmark, Kgs. Lyngby, DK-2800, Denmark}
\alsoaffiliation{Center for Free-Electron Laser Science CFEL,
Deutsches Elektronen-Synchrotron DESY, Notkestr. 85
22607 Hamburg, Germany}
\email{soco@kemi.dtu.dk}
\author{Carlo Callegari}
\affiliation[Elettra]
{Elettra - Sincrotrone Trieste S.C.p.A., Basovizza,
34149, Trieste, Italy}
\author{Russell S. Minns}
\affiliation[Soton]
{School of Chemistry and Chemical Engineering, University of Southampton, Southampton, SO171BJ, United Kingdom}
\email{r.s.minns@soton.ac.uk}
\author{Michele Di Fraia}
\affiliation[CNRIOM]
{CNR - Istituto Officina dei Materiali (IOM), Basovizza, 34149, Trieste, Italy}
\alsoaffiliation
{Elettra - Sincrotrone Trieste S.C.p.A., Basovizza,
34149, Trieste, Italy}
\email{difraia@iom.cnr.it}

\title[An \textsf{achemso} demo]
  {Shake-down spectroscopy as state- and site-specific probe of ultrafast chemical dynamics}

\abbreviations{IR,NMR,UV}


\begin{document}

\begin{tocentry}
\includegraphics[width=1\textwidth]{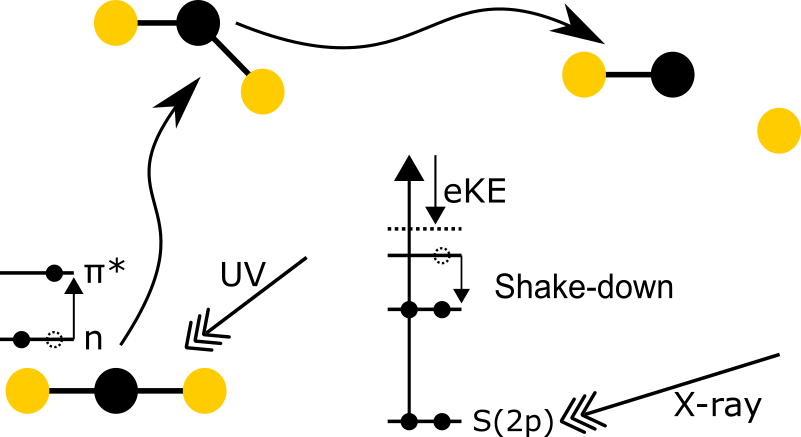}




\end{tocentry}

\newpage
\begin{abstract}
  Tracking the multifarious ultrafast electronic and structural changes occurring in a molecule during a photochemical transformation is a challenging endeavor that benefits from recent experimental and computational progress in time-resolved techniques. Measurements of valence electronic states, which provide a global picture of the bonding structure of the molecule, and core electronic states, which provide insight into the local environment, traditionally require different approaches and are often studied separately.
Here, we demonstrate that X-ray pulses from a seeded free-electron laser (FEL) enable the measurement of high-resolution, time-resolved X-ray photoelectron spectra (XPS) that capture weak satellite states resulting from shake-down processes in a valence-excited molecule. This approach effectively combines the advantages of both valence- and core-state investigations.
We applied this method to investigate photoexcited CS$_2$ molecules, where the role of internal conversion (IC) and intersystem crossing (ISC) in determining the pre-dissociation dynamics is controversial. 
We present XPS spectra from photoexcited CS$_2$, obtained at the FERMI FEL. High-resolution measurements, compared to the corresponding spectra obtained from accurate multireference quantum chemical calculations, reveal that shake-down satellite channels are highly sensitive to both valence electronic and geometric changes. Previous studies of the pre-dissociation dynamics have led to uncertain assignments of the branching between singlet and triplet excited states. We derive a propensity rule that demonstrates the spin-selectivity of the shake-downs. This selectivity allows us to unequivocally assign contributions from the bright and dark singlet excited states, with populations tracked along the pre-dissociation dynamic pathway.
   
\end{abstract}

\section{Introduction}\label{sec1}
Photochemical dynamics involves the coupled motion of electrons and nuclei that rearrange on ultrafast timescales.
Even for a simple linear triatomic molecule, measuring and understanding the reaction mechanisms upon excitation is challenging. 
The ability to spectroscopically track photochemical processes occurring in a molecule on a femtosecond timescale is based on: the availability of suitable light sources, the sensitivity of experimental methods to the electronic and geometrical changes, and accurate theoretical protocols to interpret the results.

Among the various experimental probes available~\cite{Rolles23},
photoelectron spectroscopies are particularly powerful and considered universal, since all states can in principle be ionized. Valence-shell photoelectron spectroscopy, where the binding energies of the outer (bonding) electrons are measured, is sensitive both to changes in the electronic character of the molecular states and to nuclear dynamics through changes in vibrational overlap~\cite{Blanchet1999}.  Core-shell photoelectron spectroscopy provides atomic site-specific information with the measured binding energy of the ejected electrons reporting on the \textit{chemical environment} of the target element. Differences in binding energies reported in core-shell photoelectron spectroscopy provide chemical shifts that characterize different atomic environments and are a sensitive measure of local charge, with long established usage for solids~\cite{Egelhoff1987} and for ground-state molecules~\cite{Gelius1974}. More recently, this concept has been extended to electronically excited molecules~\cite{Vidal:2020:1,Vidal:2020:2,Mayer2022, Brausse18, Leitner18,Wernet17}, providing a sensitive probe of the change in local charge experienced by specific atoms within a molecule as it photochemically evolves.

X-ray photoelectron spectroscopy (XPS) measurements of excited-state molecules can also provide valence shell information through the observation of weaker secondary structures associated with satellite transitions such as shake-up or shake-down.\bibnote{Other processes such as shake-off are possible but we limit the discussion here for the sake of brevity.} In shake-up/down, there is a rearrangement of the valence electrons as the core electron is removed. In the electronic ground state, molecules can undergo shake-up processes in which a valence electron is excited in conjunction with the core ionization transition. In such shake-up processes, the measured kinetic energy of the core-ionized electron is reduced\textcolor{black}{, relative to that of the main photoline,} by the energy associated with the valence excitation.

\begin{figure*}[hbt]
    \centering
    \includegraphics[width=1\linewidth]{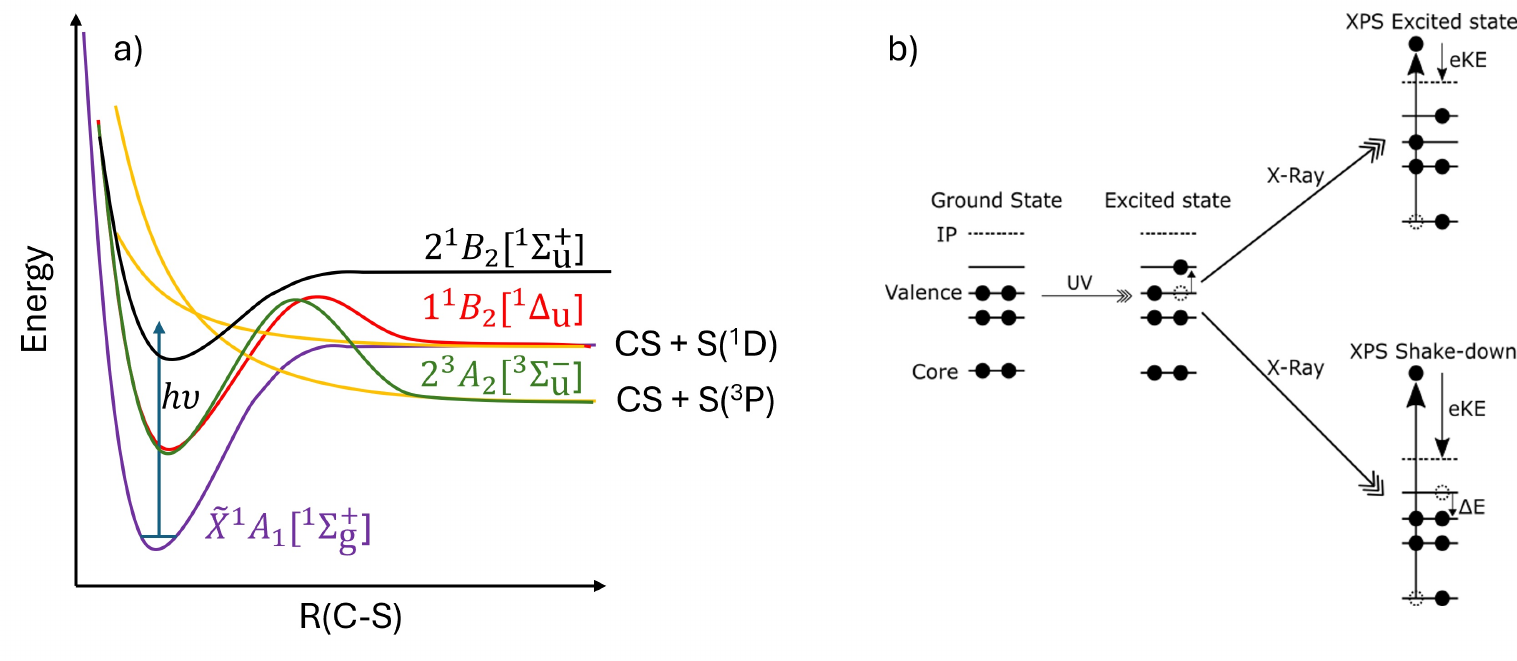}
    \caption{a) Schematic representation of the potential energy surfaces of \ce{CS2} relevant to the photodissociation process. The sketch is based on the potentials presented in \citet{Smith2018} and \citet{Gabalski2023} Excitation with a UV pump photon populates the 
    $2~{}^{1}\!B_2~[{}^{1}\!\Sigma_{\textrm{u}}^+]$ excited state. Coupling to the manifold of singlet and triplet excited states leads to cleavage of one C-S bond and the formation of CS and S in either the $^1$D or $^3$P state. Energy level representations of the photoionization processes are shown in panel b). A UV pump populates a valence excited state. Core ionization with an X-ray leads to peaks in the spectrum associated with ionization of the excited state, and to a secondary peak that has a higher electron kinetic energy (eKE) due to the relaxation of the valence excited state during the ionization process. As the molecular structure evolves on the excited state potential energy surface, the energy gap, $\Delta$E, between the electronically excited and the ground state changes, leading to time-dependent shifts in the measured kinetic energy of the electrons associated with shake-down. }
    \label{fig:Schematic}
\end{figure*}

Upon core ionization of electronically excited states, the inverse process, shake-down, is possible~\cite{Schulz2005}. In shake-down, the populated valence excited state relaxes to a lower energy configuration, increasing the energy of the outgoing photoelectron by the corresponding amount (see sketch in panel \ref{fig:Schematic}(b)). The observation of shake-down transitions from laser excited molecules and those undergoing dynamics remains largely unexplored. To observe such shake-down processes in a time-resolved experiment, highly stable and ultrafast X-ray pulses, combined with highly sensitive electron detection are required. By performing high-resolution XPS measurements with the seeded FEL FERMI we demonstrate the sensitivity of the XPS technique and the ability to measure shake-down satellite transitions, through an investigation of the photodissociation of CS$_2$. A schematic representation of the chemical dynamics and the photoionization processes is presented in Figs.~\ref{fig:Schematic}(a) and ~\ref{fig:Schematic}(b) respectively; throughout the manuscript we denote the electronic states with $C_{2\text{v}}[D_{\infty\text{h}}]$ labels\bibnote{As pointed out in footnote 1 of Ref.~\citenum{Townsend2006}, there is some inconsistency in the notation used in the existing literature for the state labels of CS$_{2}$. Therefore, we will refrain from using the adiabatic state labels or the standard empirical (tilde) notation for the valence excited states.}.
The measurements highlight the sensitivity and selectivity of the shake-down transitions to a particular spin multiplicity and allow us to spectrally isolate the singlet-state dynamics of the system.

Previous measurements of CS$_2$ have shown that despite its structural simplicity, the dynamics of UV photolysis are complex, involving coupled nuclear and electronic motion, and two dissociation channels. Excitation around 6 eV ($\sim$200~nm) leads to population of the 
$2~{}^{1}\!B_2~[{}^{1}\Sigma_{\textrm{u}}^+]$ electronically excited state. Subsequent structural changes lead to the formation of a bent and stretched structure (as seen in scattering and Coulomb explosion measurements~\cite{Razmus2022,Unwin2023,Amini2019,Gabalski2022}) that facilitates internal conversion (IC) and intersystem crossing (ISC) processes, and the eventual formation of CS fragments in conjunction with a S atom. The S atom can be formed in either the $^1$D configuration in a 
spin-allowed process, or the $^3$P configuration in the 
spin-forbidden process. Repeated experiments have shown that, despite the sub-picosecond ultrafast nature of the dissociation process, the spin-forbidden product dominates the yield. 
While this general picture is accepted, questions remain about which interactions control the branching between the singlet and triplet dissociation pathways. Conflicting assignment of existing photoelectron spectra also mean that it is unclear whether the triplet states of \ce{CS2} have any appreciable lifetime, or the dissociation is so rapid that the formation of products occurs before any population can be detected~\cite{Smith2018,Warne2021,Karashima2021}.

Time-resolved XPS measurements of excited CS$_2$ were recently reported~\cite{Gabalski2023}. By probing the dynamics at the S 2p edge by means of SASE FEL pulses, \citet{Gabalski2023} were able to assign the main X-ray spectroscopic features to the dissociation products and follow the dynamics with reported time-scales in agreement with previous valence studies.
In their paper, Gabalski et al. showcased the potential of the TR-XPS technique, and made an initial attempt to track pre-dissociation dynamics in excited CS$_2$ molecules by means of core level ionization. Our use of a seeded FEL scheme, over a SASE FEL, offers a superior energy resolution and a negligible timing jitter, allowing one to distinguish weaker secondary photoemission processes and to capture the finer details of the dynamics. 
Utilizing the highly stable output of the two-stages seeded FEL at FERMI we obtain the valence and core spectrum in a single session, including the satellite states associated with shake-down processes. The satellite states, in particular, provide a sensitive and selective probe of the electronic and geometric structure changes that occur over the course of a photochemical reaction.

\section{Results and discussion}\label{sec2}
\begin{figure*}[hbt]
    \centering
    \includegraphics[width=\linewidth]{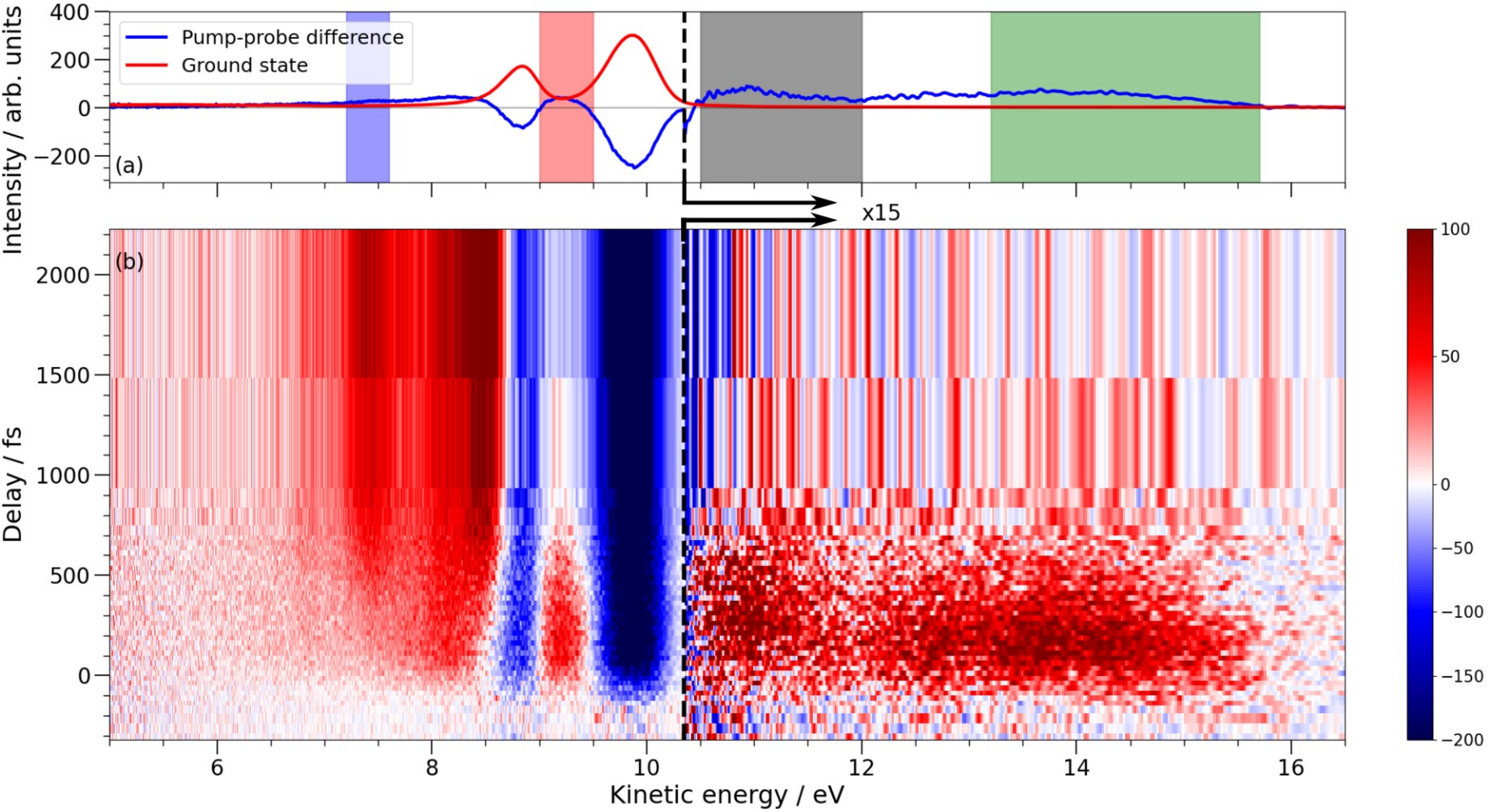}
    \caption{An overview of the time-resolved difference X-ray photoelectron spectrum of CS$_2$ (b) obtained following 200~nm excitation, and ionization with a 179.9~eV probe with a comparison (a) of the ground state (red) spectrum scaled by 0.1 and the 0 – 600~fs pump-probe difference spectrum (blue). The colour of the shaded regions corresponds with the integrated intensity of the same colour within Fig.~\ref{fig:integrated intensity}.}
    \label{fig:XPS}
\end{figure*}

In Fig.~\ref{fig:XPS} we provide an overview of the experimental XPS data obtained following 6.2 eV (200 nm) excitation, and ionization with a 179.9 eV probe.  The figure provides a map of the differential (pump-on minus pump-off) photoelectron intensity as a function of electron kinetic energy and time. \textcolor{black}{The blue regions around 8.8~eV and 10~eV are due to depletion of the signals associated with the ionization of the S 2p$_{1/2}$ and S 2p$_{3/2}$ orbitals of the ground state of the \ce{CS2} molecule. Their kinetic energies correspond to binding energies of 171.08 eV and 169.81 eV, respectively \cite{Hedin2009}}. As the ground-state signal depletes, we see new, transient signals appearing at both higher and lower kinetic energies that decay on a sub-ps timescale. At longer delays, signals at even lower kinetic energy grow and reach a constant level within 2~ps of excitation, indicating that these relate to the CS and S dissociation products. \textcolor{black}{The signals in the lower kinetic energy region show the same trends as those observed by \citet{Gabalski2023} but with greater energy resolution and signal-to-noise ratio. Signals at energies below 10.5~eV are therefore assigned as XPS photolines associated with the various reaction intermediates and product states created.} At kinetic energies greater than 10.5~eV we see much weaker transient signals that cannot be assigned to direct X-ray ionization: these are related to shake-down transitions. In this higher kinetic energy range the signal is initially localised to energies $\lesssim$ 16 eV and its distribution shifts towards lower kinetic energies as a function of time.

\begin{figure*}[hbt]
    \centering
    \includegraphics[width=\linewidth]{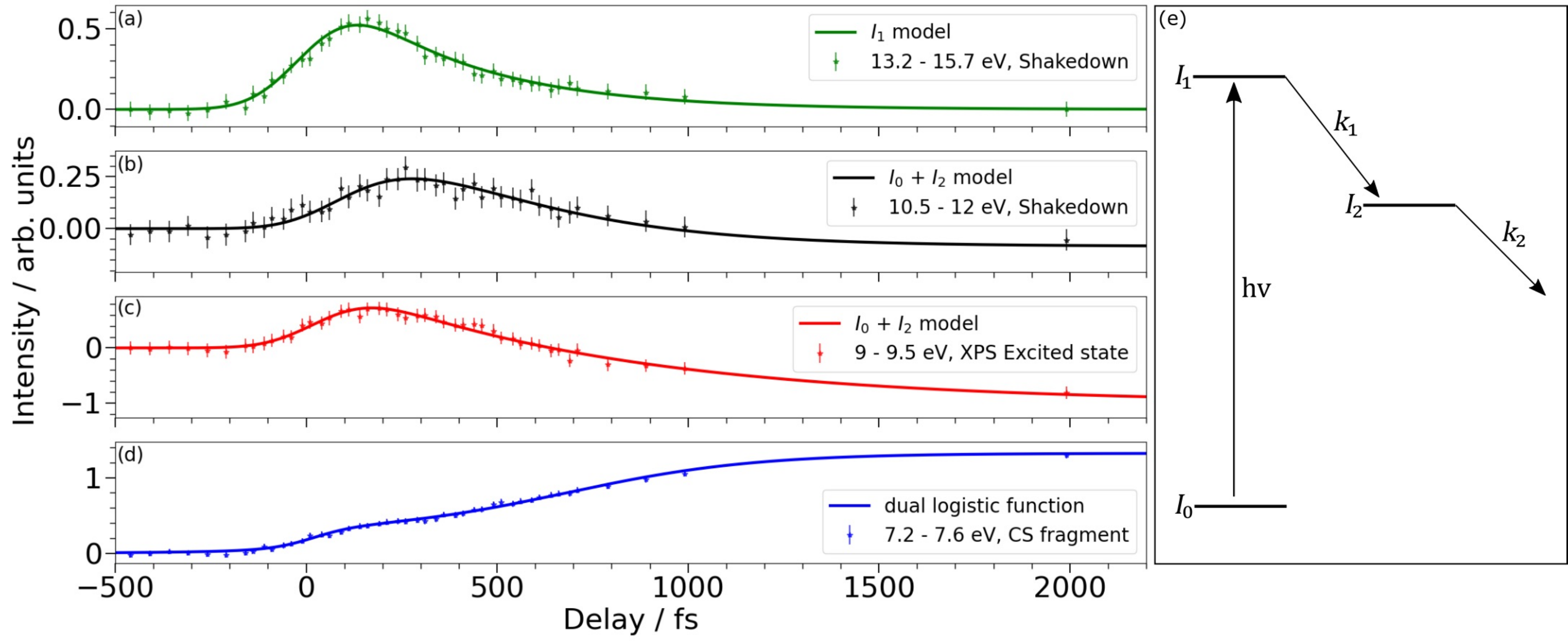}
    \caption{Integrated intensity of the photoelectron bands observed in the energy ranges 13.2-15.7 eV (a), 10.5-12.0 eV (b), 9.0-9.5 eV (c), and 7.2-7.6 eV (d). e) Schematic representation of the kinetic model used to fit the experimental data. $I_0$, $I_1$ and $I_2$ represent changes in state population associated with the laser excitation step, $h\nu$, and subsequent exponential decays of the populations in the two excited states defined by the rate constants, $k_n$. The equations derived from the model and used to fit the data are given in SI section~S1.5.} 
    \label{fig:integrated intensity}
\end{figure*}

To explore the transient nature of the signals appearing in various regions of the spectrum in a more quantitative manner, we plot the integrated intensity over a number of kinetic energy ranges in Fig.~\ref{fig:integrated intensity}.  Panels (a) and (b) show the intensity profile of the shake-down transitions observed in the spectrum. We separately integrate over the 13.2--15.7 eV (a) and 
10.5--12.0 eV (b) regions which show quite different temporal dynamics. The profiles show that the photoelectron intensity is initially localised in the higher kinetic energy range, (a). As this intensity decays, we observe a commensurate increase in intensity in (b), thereby indicating a flow of population between states. To extract lifetimes from the data we fit the intensity profiles to a sequential kinetic model as schematically represented in Fig.~\ref{fig:integrated intensity}(e) (see SI 
section~S1.5 
for more details and for the equations used), where excitation leads to population of the region (a) which subsequently decays into region (b). The fits are plotted as solid lines in Fig.~\ref{fig:integrated intensity} (a) and (b) and provide exponential time constants of 345~$\pm$~12~fs and 167~$\pm$~15~fs, respectively. These values match those previously obtained in valence photoelectron spectroscopy measurements of \ce{CS2} for the excited states populated en route to dissociation~\cite{Smith2018,Karashima2021}. At FERMI we have the advantage of being able to measure the XPS and valence photoelectron spectrum within the same instrument and we obtain equivalent time constants in the time-resolved valence photoelectron spectrum (\textcolor{black}{see Figures S6 and S7} in SI 
section~S2 
for details).

Figure \ref{fig:integrated intensity}(c) shows the intensity profile for the transient increase in signal observed in the 9.0-9.5 eV range, that is, between the two spin-orbit split ground state depletion peaks. The transient enhancement observed in (c) is fit to an equation including ground state depletion and transient population of an excited state. The resultant fit shows that the decay of the transient excited state signal in panel (c) has \textcolor{black}{the same} temporal profile as the signals plotted in panels (a) and (b) combined. The common intensity profile indicates that the two energy regions, 9.0-9.5 and 10.5-15.7 eV, report on the same excited state populations but with the shake-down region showing greater spectral separation between components.

Figure \ref{fig:integrated intensity}(d) shows the signal associated with the formation of the CS dissociation product in the 7.2-7.6 eV range.
The CS product signal in panel (d), shows two time separated rises. We fit the rise to a sum of two logistic functions to characterise these rises and obtain an appearance time, which defines the time at which 50\% of the maximum intensity is obtained, as used in many time-resolved ion yield experiments. The logistic functions return appearance times of 20~$\pm$~10~fs and 685~$\pm$~23~fs with relative amplitudes of 0.25 to 1 for the fast and slow channels, respectively. The 685~fs appearance time matches the decay associated with the excited state signals observed in the shake-down and XPS regions of the spectrum.

\begin{figure*}[hbt]
    \centering
    \includegraphics[width=\linewidth]{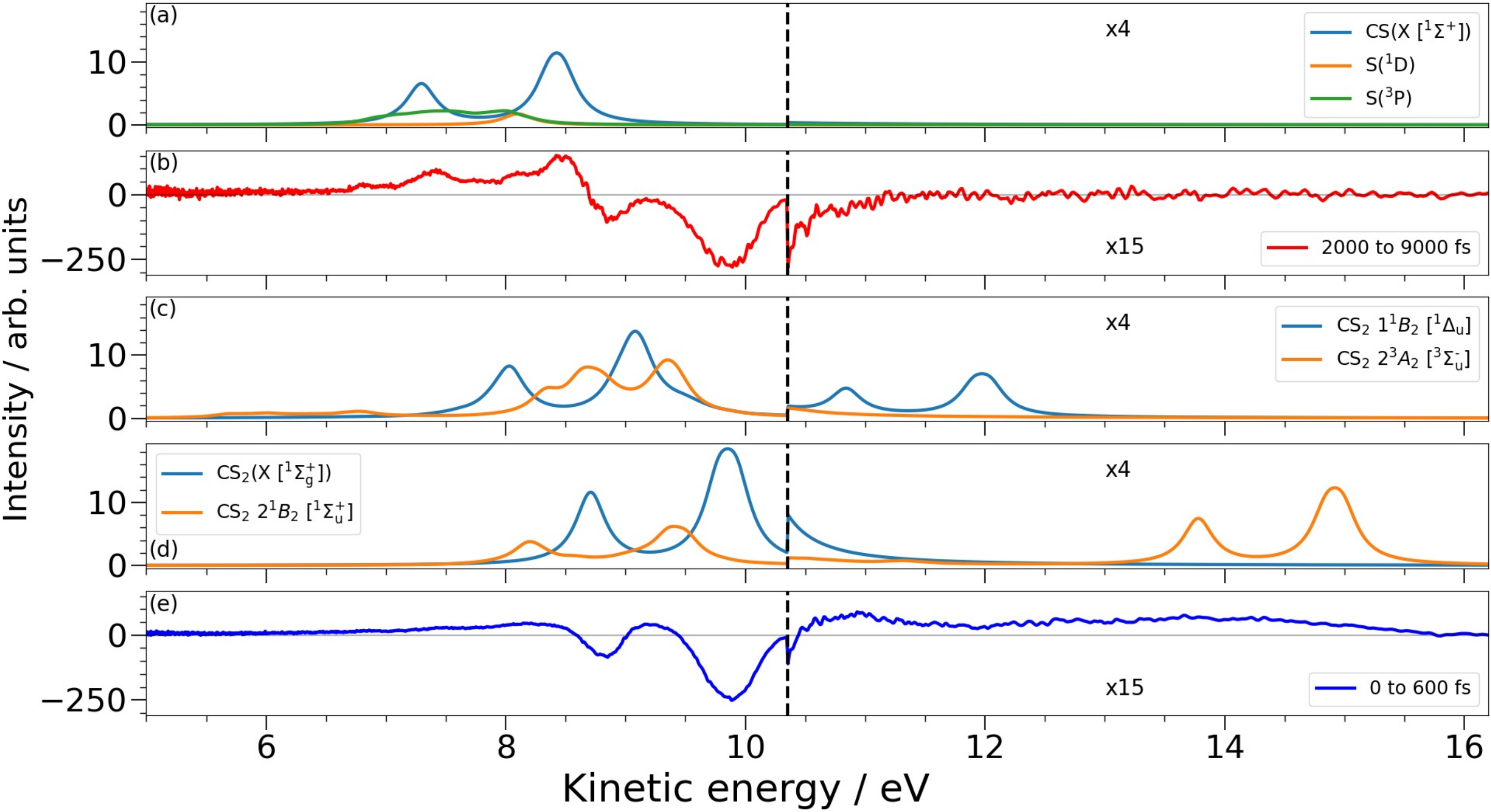}
    \caption{Comparison of theoretically calculated XPS spectra for the dissociation products (a), and the electronic states of \ce{CS2} (c) and (d) with the experimental spectra obtained at early (e) and late (b) delays. \textcolor{black}{All calculated spectra have been shifted by 0.76 eV and broadened with Lorentzian functions with a full width at half maximum of 0.3 eV.}}
    \label{fig:Exp_Theory_comp}
\end{figure*}

To assign the peaks observed in the experiment, we computed the X-ray photoelectron spectrum of CS$_2$ in the ground state and in all energetically accessible electronically excited states. The calculations are performed at the equilibrium geometry associated with that particular electronic state (reported in section~S3.1),
with the results of all calculations and further details provided in the SI, see sections~S3.3
to S3.5.
\textcolor{black}{The theoretical XPS spectra of the initially populated valence excited state, as well as the identified IC- and ISC-accessible excited states are plotted in Fig.~\ref{fig:Exp_Theory_comp}.  Below the calculations are time-averaged experimental spectra obtained at early delays, where the various electronically excited states are populated, and at late delays where products have been formed.}The calculated kinetic energies for all states are shifted by 0.76~eV to match the ground-state energies obtained in the experiment. No other scaling or corrections are applied.  

Calculations of the \textcolor{black}{spectra for ground state, $^1\Sigma_{\textrm{g}}^+$, and the initially excited state, 
$2~{}^1\!B_2~[^1\Sigma_{\textrm{u}}^+]$, Fig.~\ref{fig:Exp_Theory_comp} (d), show that upon excitation the photoelectron kinetic energy associated with direct ionization of the electronically excited state shifts towards lower kinetic energy. }This is explained by the $\pi^*\leftarrow n$ character of the transition that reduces the local electron density on the S atom, thereby increasing the binding energy of the S 2p electrons. The shift is relatively small, on the order of 0.5~eV and leads to the appearance of the transient enhancement at kinetic energies between the two spin-orbit coupled peaks of the \textcolor{black}{electronic} ground state. 

Ionization of the S 2p electron also gives rise to weaker signals at kinetic energies around 14--16 eV. These peaks are related to shake-down transitions where the valence hole is filled upon core ionization, leading to an increased kinetic energy of the outgoing electron. The peaks show an increase in kinetic energy relative to the core ionization peaks with the energy gained by the electron being controlled by the energy gap between the valence excited state and the ground electronic state. The calculations of the ${}^1\!B_2~[^1\Sigma_{\textrm{u}}^+]$ electronic state are performed at the bent equilibrium geometry of the state, giving rise to the two sharp peaks associated with the two spin-orbit coupled states. We stress that in the experiment the molecule is known to asymmetrically stretch and bend such that a single dominant geometry is unlikely. As the energy spacing between the linear, ground state, and the bent ${}^1\!B_2~[^1\Sigma_{\textrm{u}}^+]$ excited state is heavily dependent on the structure, the energy gained upon shake-down is strongly geometry dependent. This leads to a smearing out of the photoelectron signal into the broad band in the 13--16~eV range observed in the experiment, which reflects the range of geometries probed at each delay.

Subsequent IC and ISC processes could lead to a number of energetically accessible states. To characterise which state or states may correlate with the shake-down signal seen in the 10.5--12.0~eV range, we calculate the expected XPS spectrum of the four lowest singlet and triplet excited states. The results of these calculations are plotted in \textcolor{black}{Figures S18 and S19} in the SI and highlight that only the  ${}^1\!B_2~[^1\Delta_{\textrm{u}}]$ state produces a signal in this region, and the calculated XPS spectrum of this state is shown in Fig.~\ref{fig:Exp_Theory_comp} (c). The kinetic energy of the direct XPS signal has a very similar shape to that of the initially populated ${}^1\!B_2~[^1\Sigma_{\textrm{u}}^+]$ with a shift of only a few tenths of an eV expected due to the limited change in local charge density on the S atoms ionized. 
The shift in the shake-down region is, however, very large, on the order of 3 eV between the two states. This large shift can be understood in relation to the schematic potentials plotted in Fig.~\ref{fig:Schematic} (a) where the energy gap between the ${}^1\!B_2~[^1\Sigma_{\textrm{u}}^+]$ and the ground state is much larger than that between the ${}^1\!B_2~[^1\!\Delta_{\textrm{u}}]$ state 
and ground state. The shifts predicted by theory match the experiment exceptionally well, suggesting that we can detect the different electronic states populated via the shake-down processes sensitively and selectively. 

We also plot the expected spectrum for the 
${}^3\!A_2~[^3\Sigma^{-}_{\textrm{u}}]$ 
state in Fig.~\ref{fig:Exp_Theory_comp} (c) as a characteristic spectrum for any triplet states that may be populated. Spectra for the other energetically accessible triplet states are provided in the SI \textcolor{black}{Figure S18} and have a very similar appearance that overlaps strongly with the singlet state XPS signals. \textcolor{black}{This indicates that we cannot use the direct XPS signals to differentiate between the accessible electronically excited states.} Critically, the triplet state spectra show no evidence of any shake-down transitions. 

Because there is little exploration of shake-down processes for electronically excited molecules, some considerations are appropriate.
%
\textcolor{black}{
In the sudden approximation, a monopole selection rule applies, such that the probability of a given transition is related to the squared overlap of the initial state with the ionized orbital removed and the final ionic 
state -- that is, to the squared norm of the relative Dyson orbital.
In the present case, the initial state is characterized by an excited $\varphi\varphi^\ast$ electron pair,
either singlet coupled (after the initial excitation) or triplet coupled (after ISC). 
 The symbols $\varphi$
 and $\varphi^*$ refer, respectively, to occupied and virtual molecular orbitals in the closed-shell ground state.
The shake-down state is characterized by an unpaired core electron in orbital $\varphi_c$ (and a $\varphi\bar{\varphi}$ singlet electron pair). 
When computing the overlap, the role of the other ``passive'' electrons (i.e., the electrons not involved in the spin coupling of the final state) will give a similar contribution to all states, and the overlap will therefore be dominated by that of the 
three active electrons (the one unpaired core electron and the two electrons involved in the initial valence excitation).
We show in Sect.~\ref{sec5}, that only molecules that at the time of the core ionization event were in a singlet valence-excited state give rise to non-vanishing shake-down signals.
In the primary ionization of the (singlet or triplet) valence excited state, the same electron 
pair
(with the same spin coupling) appears in the final state, giving similar 
intensities for both couplings.
}

As CS$_2$ has a singlet electronic ground state, we therefore expect to observe predominantly shake-down transitions originating from singlet excited states and not from triplet excited states. 
This general picture is seen in the shake-up spectrum of ground state, core-ionized CS$_2$ of \citet{Wang2001} where the shake-up intensity of the triplet states is approximately an order of magnitude lower than that of the equivalent singlet state. 
This appears to be borne out by the theoretical predictions (see \textcolor{black}{Sect.~\ref{sec5}}) and the experiment. 
While further study of shake-down is required, based on these arguments, we assign all of the signal in the
shake-down region to singlet state populations.

Turning to longer delays and the formation of products, we present calculated spectra associated with the \ce{CS}, and \ce{S}~(${}^1$D and ${}^3$P) fragments in Fig.~\ref{fig:Exp_Theory_comp}(a), and in Fig.~\ref{fig:Exp_Theory_comp}(b) we show the experimental spectrum for late delays. The computed spectra agree extremely well with the experimental observation, reproducing the peak positions and relative intensities very well. We note that at early times there could be significant overlap between the signal associated with excited states of bound \ce{CS2} and that of the \ce{CS} fragment peak appearing at kinetic energies around 8.5 eV. We therefore use the weaker of the two \ce{CS} peaks at 7.5~eV to monitor the product appearance times, Fig.~\ref{fig:integrated intensity}(d), as this appears to be free from any such contamination. 

The combined experimental and theoretical results provide the following picture of the dynamics. Excitation into the ${}^1\!B_2~[^1\Sigma_{\textrm{u}}^+]$ state initiates bending and stretching of the nuclear framework. A minor amount of this population undergoes ballistic dissociation leading to formation of \ce{CS} and \ce{S}(${}^1$D). The remaining population internally converts to the ${}^1\!B_2~[{}^1\Delta_{\textrm{u}}]$ state on a 350~fs timescale. The ${}^1\!B_2~[{}^1\Delta_{\textrm{u}}]$ population decays on a 160 fs timescale leading to CS + S fragments. The ability to directly assign specific excited states in the shake-down spectrum allows us to settle the controversy in the literature and assign the singlet intermediate state populated to the ${}^1\!B_2~[{}^1\Delta_{\textrm{u}}]$. While spectral congestion means we cannot directly assign the dissociation product to either the singlet or triplet channel, previous studies show that the overriding majority of this decay should lead to the \ce{S}($^3$P) product. The measurements show that the decay of the excited singlet state signal observed in the shake-down region of the spectrum is on the same timescale as the formation of the \ce{CS} products. One explanation could be that an unobserved triplet state intermediate is also populated and decays on an indistinguishable timescale. We find this suggestion unlikely given the dissociation limit for the triplet channel is approximately 1.1 eV lower than that for the singlet channel, and the shallower energy minimum of the triplet states compared to the singlet counterparts~\cite{Smith2018}. We therefore conclude that any \ce{CS2} triplet state population is extremely short-lived, and that dissociation proceeds ballistically once intersystem crossing has occurred.

\textcolor{black}{The proposed reaction mechanism can be summarised in the following scheme:
\schemestart 
\ce{CS2(X) + h$\nu$}
\arrow(aa--){->}[-90]
\ce{CS2(${}^1\!B_2~[^1\Sigma_{\textrm{u}}^+]$)}
\arrow(aa--){->[Diss]}
\ce{CS + S($^1$D)}
\arrow(@aa--){->[*{0}IC]}[-90]
\ce{CS2(${}^1\!B_2~[{}^1\Delta_{\textrm{u}}]$)}
\arrow{->[ISC][Diss]}
\ce{CS + S($^3$P)}
\label{Sch:CS2}
\schemestop\\
\\
We highlight here that the proposed mechanism is based on the observation of the shake-down region of the excited state spectrum and of the product state XPS. Contrary to the direct XPS signals, where different electronic states produce highly overlapped peaks in the spectrum, analysis of the shake-down region allows us to exclusively monitor the singlet state population and distinguish the character of the populated excited electronic states. By monitoring the singlet state population and the product formation channels, we can draw reasonable conclusion about the triplet states despite them not being observed directly. In fact, due to the proposed mechanism it appears likely that the instantaneous population of the triplet states of bound \ce{CS2} may be negligible such that probes that are highly sensitive to the triplet state population may struggle to detect these states.  }

\section{Conclusion}\label{sec3}
We report the first time-resolved observation of shake-down transitions from laser-excited \ce{CS2} molecules probed above the S 2p edge with FEL radiation.
In particular, upon excitation with 200 nm pump laser, we observed the shake-down signals associated with the $2~{}^1\!B_2~[^1\Sigma_{\textrm{u}}^+]$ and the $1~{}^1\!B_2~[{}^1\Delta_{\textrm{u}}]$ excited states. These peaks are significantly shifted from one another when compared to those observed in the direct XPS. The large shifts seen in the position of the shake-down peaks as the molecules change geometry and electronic state, allow us to unequivocally assign the states populated during the pre-dissociation process and characterize the internal conversion from the bright excited state into the dark excited state. 
The delay in the appearance of the latter with respect to the former gives an indication of the time-scale for the internal conversion to occur in excited \ce{CS2} molecules. 

The results demonstrate the detailed and state-specific information that can be extracted from the shake-down process, over and above what could be extracted from the primary ionization alone. The large shifts in electron kinetic energy, related to the energy spacing between the valence excited state and the electronic ground state, provide a distinct observable that can be definitively assigned to specific states in the excited state manifold. 

The observation of shake-down transitions opens a new spectroscopic window into the electronic structure changes that occur during a complex chemical reaction. The characteristics of the ground and excited states that lead to observable shake-down signals is currently unexplored, but lessons can be learned from static measurements of the shake-up process which will share the same selection and propensity rules. In principle, provided the transitions meet the symmetry selection rules, all states can be coupled. The cross-section of such transitions is, however, less clear and requires significant further exploration. The propensity of an allowed transition will significantly depend on the orbital overlap and the effective strength of the monopole/dipole with the states involved. We note here that the highest occupied molecular orbital that defines the start (end) point of the pump (shake-down) transition, in our particular study on \ce{CS2}, is characterised as a lone pair orbital localised on the S atom. As such, this will have a strong interaction with the monopole derived from core ionization that is localised again on the S atom. Therefore, one might expect that ionization from a C core may lead to a lower propensity for shake-down. Such sensitivity, if demonstrated, provides a highly localised probe of the valence electronic state character which could enable tracking of the changing character of electronically excited states in much larger molecules.

\section{Methods}\label{sec4}
\subsection{Experimental Details}
The experiment has been performed at the Low Density Matter (LDM) beamline~\cite{Svetina2015} of the Free Electron Laser FERMI~\cite{Allaria2015} in Trieste (Italy). The soft x-ray FEL probe pulse is generated in the FEL-2 machine \cite{Allaria2013}, set to produce pulses at harmonic 12 of the seed laser wavelength 248 nm in the first stage, and harmonic 3 of the resulting 20.67 nm pulse (6.89 nm) in the second stage, at 50 Hz repetition rate; metal foil filters are available along the photon transport line to alter the balance of the two pulses. In particular, a palladium metal filter (Pd 100 nm) was used to abate the first-stage radiation. 
The FEL spot size was set to 50 $\mu$m (FWHM) and the pulse duration was estimated to be approximately 30 fs. 

The pump laser setup is based on the availability of
IR pulses, referred to as SLU (Seed Laser for Users) generated by a Ti:Sapphire amplifier having the same laser oscillator as that used to drive the OPA generating the seed pulses \cite{Cinquegrana2021}. IR SLU pulses are optically transported to an optical table. The basic optical layout on this table is described in \cite{Finetti2017}; it has subsequently been upgraded to include fourth harmonic generation (via sequential second-harmonic generation and sum-frequency generation). For this experiment, the UV excitation pulse was generated with a central wavelength of 199.72 nm and bandwidth of 0.86 nm FWHM (6.21 $\pm$ 0.03 eV); focus size 150 $\mu$m (FWHM); pulse duration 110 fs (FWHM); 25 Hz repetition rate. Up to 5 $\mu$J of UV light could be generated, greatly exceeding the needs of the experiment (most measurements were done at 0.2 $\mu$J, by reducing the intensity of the input IR).
The SLU repetition rate was set to half of the FEL repetition rate to have FEL only shots and FEL + SLU shots to generate differential spectra in post-acquisition.   
The pump–probe instrument response function (IRF) was estimated by monitoring the XPS S 2p ground state depletion signals and resulted in an IRF of 108 $\pm$ 5.8 fs (for details, see the supporting information).  
When needed, the second stage of the FEL was ``turned off'' and the first stage alone at harmonic 5 of a seed wavelength at 261.1 nm was used to find and periodically check the temporal and spatial overlap of the FEL and SLU pulses. The resulting wavelength, 52.22 nm, was chosen because it coincides with the 1s5p $\leftarrow$ 1s$^2$ resonance of helium. The spatial and temporal optimization was done by monitoring the UV-induced ionization of helium atoms resonantly excited by the FEL pulse. 

The basic layout of the endstation described in \cite{Lyamayev2013} does not include the Magnetic Bottle Electron Spectrometer (MBES), which became available as a later upgrade \cite{Squibb2018} and was the spectrometer of choice in the present experiment, with the axis of the MBES oriented vertically. For the TR-XPS maps of Fig.~\ref{fig:XPS} a retardation voltage of 4 V was applied to increase the resolution in the region of the S 2p XPS peak.
A mixture of 0.14 bar CS${_2}$ in 2 bar He was expanded into vacuum with a commercial pulsed valve (Parker Series 9, orifice aperture 250 $\mu$m) operated at the FEL repetition rate, 50 Hz, and nominal opening time 110 $\mu$s, in the endstation's source chamber; a supersonic jet is formed along the horizontal long axis of the endstation, and passes through a conical skimmer (Beam Dynamics model 76.2, 3 mm diameter) into a differential pumping chamber, where it is further defined by a fixed-diameter iris (1.5 mm) and a set of piezoelectrically-operated vertical slits (Piezosystem Jena PZS 3) before entering the detector chamber, where it crosses perpendicularly the FEL beam, the latter propagating along the horizontal short axis of the endstation. The SLU beam is sent into the detector chamber quasi-collinearly with the FEL (4$^\circ$ downward tilt).
The absorption spectrum of CS$_2$ measured by
\citet{Hemley1983} suggests that the
transitions excited by the SLU originate mainly from
the vibrationally unexcited ground state, but with a
small contribution from the level having one quantum
of the bending mode. The upper levels populated
will contain several quanta in the symmetric stretching
and/or bending vibrational modes.

The calibration details of the spectrometer are outlined in the Supplementary Information \textcolor{black}{in particular Figures S2 and S5}.

\subsection{Computational Details}
All initial and final states have been computed with the multi-state restricted active space perturbation theory to second-order (MS-RASPT2) method, as implemented in OpenMolcas \cite{Openmolcas-JCTC-2023}. $C_{\textrm{2v}}$ point
group symmetry was applied for the CS$_2$ and CS molecules. Specifically, the linear CS$_2$ and CS molecules were aligned along the y axis, while the bent CS$_{2}$ molecules were oriented so that the yz plane constitutes the molecular plane, and the C$_{2}$ rotation axis is oriented along the z axis. The $D_{\textrm{2h}}$ point group symmetry was applied for the S atom. All calculations were performed with the ANO-RCC-VTZP base set \cite{roos2004main}. Scalar relativistic effects were taken into account via the 
Douglas-Kroll-Hess
Hamiltonian~\cite{nakajima2012douglas},
and spin-orbit coupling via the a posteriori
addition of the spin–orbit part of the DKH Hamiltonian as an effective mean-field one-electron operator~\cite{malmqvist2002restricted}.

The orbital configuration of the ground state, including the lowest virtual orbitals, is reported in
Table~S7
in SI.
The active space for CS$_2$ was constructed with all the S 2p orbitals (symmetric 4a$_1$, 5a$_1$, and 1b$_1$, and anti-symmetric 3b$_2$, 4b$_2$, and 1a$_2$) in the RAS1 subspace, followed by the occupied valence orbitals 7a$_1$ ($\sigma$), 8a$_1$ ($\pi$), 6b$_2$ ($\sigma$), 7b$_2$ ($n$), 2a$_2$ ($n$), and 2b$_1$ ($\pi$) and virtual orbitals 9a$_1$ ($\pi^*$), 10a$_1$ ($\sigma^*$), 8b$_2$ ($\sigma^*$), and 3b$_1$ ($\pi^*$) in the RAS2 subspace. The active space for CS was formed by placing the S 2p orbitals (4a$_1$, 1b$_1$, and 1b$_2$) in the RAS1 subspace, and the occupied valence orbitals 5a$_1$ ($\sigma$), 6a$_1$ ($n$), 7a$_1$ ($\sigma$), 2b$_1$ ($\pi$), and 2b$_2$ ($\pi$) as well as the virtual orbitals 8a$_1$ ($\sigma^*$), 3b$_1$ ($\pi^*$) and 3b$_2$ ($\pi^*$) in the RAS2 subspace. Lastly, the active space for S contained the 2p orbitals (1b$_{\mathrm{1u}}$, 1b$_{\mathrm{2u}}$, and 1b$_{\mathrm{3u}}$) in the RAS1 subspace, followed by the occupied 3s (3a$_\mathrm{g}$) and 3p (2b$_{\mathrm{1u}}$, 2b$_{\mathrm{2u}}$, and 2b$_{\mathrm{3u}}$) orbitals as well as the virtual 4s (4a$_\mathrm{g}$), 3p (3b$_{\mathrm{1u}}$, 3b$_{\mathrm{2u}}$, and 3b$_{\mathrm{3u}}$), and 3d (5a$_\mathrm{g}$, 6a$_\mathrm{g}$, 1b$_{\mathrm{1g}}$, 1b$_{\mathrm{2g}}$, and 1b$_{\mathrm{3g}}$) orbitals in the RAS2 subspace. 
The RAS3 subspace was kept empty, and we allowed for a maximum of one hole in the RAS1 subspace in all cases.
The active orbitals are displayed in section~S3.2
of the SI \textcolor{black}{(Figures S8-S18)}. 

All XPS spectra were computed using the (state-specific) restricted active space state-interaction (RASSI) module \cite{RASSI-MALMQVIST}, and we note that the Dyson amplitudes were calculated 
from biorthonormally transformed orbital sets~\cite{tenorio2022photoionization},
and using the spin-orbit coupled states. The final core-ionized states of the CS$_2$ and CS molecules were obtained by state-averaging over 20 states per irreducible representation. For the S atom, we computed only one final core-ionized state per irreducible representation, except for the $A_{\mathrm{g}}$ irreducible representation, where we state-averaged over two final states. The core-ionized states were obtained by enforcing single electron occupation in the RAS1 subspace by means of the HEXS projection technique \cite{HEXS-RAS}. An imaginary shift of 0.3 a.u. was applied to all states in the second-order perturbation correction to avoid intruder states. 

The XPS spectra of CS$_{2}$ \textcolor{black}{(Figs. S18, S19)} were computed for the ground state ($\Tilde{X}~{}^{1}\!A_{1} ~[{}^{1}\Sigma^{+}_{\mathrm{g}}]$) and for 
the following singlet valence-excited states 
$1~{}^{1}\!A_{2}~[{}^{1}\Sigma^{-}_{\mathrm{u}}]$, $2~{}^{1}\!A_{2}~[{}^{1}\!\Delta_{\mathrm{u}}]$,  
$1~{}^{1}\!B_{2}~[{}^{1}\!\Delta_{\mathrm{u}}]$, 
$2~{}^{1}\!B_{2}~[{}^{1}\Sigma^{+}_{\mathrm{u}}]$, 
and triplet valence-excited states
$1~{}^{3}\!B_{2} ~[{}^{3}\Sigma^{+}_{\mathrm{g}}]$, 
$1~{}^{3}\!A_{2}~[{}^{3}\!\Delta_{\mathrm{u}}]$, 
$2~{}^{3}\!B_{2}~[{}^{3}\!\Delta_{\mathrm{u}}]$,
$2~{}^{3}\!A_{2} ~[{}^{3}\Sigma^{-}_{\mathrm{u}}]$ 
where we use the notation
$C_{2\mathrm{v}}~[D_{\infty\mathrm{h}}]$ ([$C_{\infty\mathrm{h}}$] for CS). 
The XPS of CS$_{2}^{+}$ was computed for the ground state 
$\Tilde{X}~({}^{2}\!A_{2}={}^{2}\!B_{2})~[{}^{2}\Pi_{\textrm{g}}]$, whereas
the XPS of CS was computed for the ground state $\Tilde{X}~{}^{1}\!A_{1}~[{}^{1}\Sigma^{+}_{\mathrm{g}}]$. 
Each XPS spectrum was constructed using initial and final states computed at the same (relaxed) structure.
For each initial (ground or valence-excited) state XPS we used the optimized molecular structure of that specific state.
The optimized structures are specified in the SI. The XPS of the S atom were computed for the ${}^{1}$D and ${}^{3}$P states. All resulting XPS spectra were averaged by the state degeneracy of the initial state. Lastly, the spin coupling between initial and final states has been taken into account by scaling
the squared Dyson amplitudes of the singlet-to-doublet transitions
by a factor of 2, those of doublet-to-triplet transitions by 3/2, and those of triplet-to-quartet transitions by 4/3.

\subsection{Propensity rule for the shake-down states}\label{sec5}

We propose a propensity rule stating that, as a general case, shake-down signals,
such as those we observed here, will originate \textcolor{black}{predominantly} from molecules that at the time of the core ionization
event were in a singlet state,
\textcolor{black}{whereas} the participation of triplet valence-excited states \textcolor{black}{is negligible}.

\textcolor{black}{In general, the appearance of satellite states can be discussed either with frozen or relaxed orbitals. Within the frozen orbital picture, the relaxation induced by the formation of a core hole may be described by means of single (and multiple, less intense) excitations with respect to the core-hole reference configuration. 
Alternatively, we may work directly with relaxed orbitals, which is the approach we adopt in the following.}

\textcolor{black}{The relative intensity of the photoionization event is typically approximated as the squared Dyson amplitude, see e.g. Ref.~\citenum{moitra2020jpcl}},
\begin{equation}
    I \sim 
    \sum_{\tilde{c}=c,\bar{c}}| 
    \braket{
     \Psi_{\mathrm{final}}|
     a_{\tilde{c}} |  
    \Psi_{\mathrm{initial}}}
    |^{2}~,
\end{equation}
\textcolor{black}{
where $a_{\tilde{c}}$ is the annihilation operator that removes either an $\alpha$ or a $\beta$ core electron. 
$\Psi_{\mathrm{initial}}$ and $\Psi_{\mathrm{final}}$ are the wave functions of the initial and final electronic states.
In the following, standard barred notation will be used when referring to the $\beta$ electrons. Hence, $\tilde{c} \in \{c, \bar{c} \}$. 
The shake-down state amounts to a primary core ionization of the closed-shell ground state (e.g., the removal of a $\beta$ core electron), and it may be written as 
$|\varphi_{c}\cdots\varphi \bar{\varphi}\rangle$.}

\textcolor{black}{
The initial state may be a singlet or triplet valence-excited state, and may be written as}
\begin{equation}
\ket{\varphi_c\bar{\varphi}_c\cdots(\varphi\varphi^{*}
    )^\textrm{S}} \quad \textrm{or} \quad \ket{\varphi_c\bar{\varphi}_c \cdots (\varphi\varphi^{*})^\textrm{T}}
\end{equation}
\textcolor{black}{where $(\varphi\varphi^*)^{\mathrm{S,T}}$ denotes the spin symmetry of the electrons involved in the initial valence excitation (singlet or triplet). Specifically,}
\begin{equation}
(\varphi\varphi^{*})^\textrm{S} = \tfrac{1}{\sqrt{2}} (\varphi \bar{\varphi}^{*} - \bar{\varphi}\varphi^{*}) ~,
\end{equation}
\begin{equation}
    (\varphi\varphi^{*})^\textrm{T} = \begin{cases}
        \varphi \varphi^{*} \\
        \frac{1}{\sqrt{2}} (\varphi \bar{\varphi}^{*} + \bar{\varphi}\varphi^{*}) \\
        \bar{\varphi} \bar{\varphi}^* ~.
    \end{cases} 
\end{equation}

\textcolor{black}{Hence, the relative intensities of the shake-down signals are found by computing}
\begin{equation}
    I_{\mathrm{S}} = 
    \sum_{\tilde{c}=c,\bar{c}}| \bra{\varphi_c \cdots\varphi \bar{\varphi}} a_{\tilde{c}} \ket{\varphi_c \bar{\varphi}_c \cdots (\varphi\varphi^{*})^\textrm{S}} |^{2} ~,
\end{equation}
\begin{equation}
    I_{\mathrm{T}} = 
\sum_{\tilde{c}=c,\bar{c}}| \bra{\varphi_c \cdots\varphi \bar{\varphi}} a_{\tilde{c}} \ket{\varphi_c \bar{\varphi}_c \cdots (\varphi\varphi^{*})^\textrm{T}} |^{2} ~.
\end{equation}
\textcolor{black}{Note that the initial $\ket{\varphi_c \bar{\varphi}_c \cdots (\varphi\varphi^{*})^\textrm{S,T}}$ and final $\ket{\varphi_c \cdots\varphi \bar{\varphi}}$ states are constructed from different  orbital sets (hence non-orthogonal to each other), 
meaning that the overlaps above are computed as the determinant of the overlap matrix between the two orbital sets. 
Below we will omit explicit indication of the ``passive'' electrons (i.e., the electrons not involved in the spin coupling of the final state),
and write the relative intensities as}
\begin{equation}
    I_{\mathrm{S},\tilde{c}} = | \bra{\varphi_c \varphi \bar{\varphi}} a_{\tilde{c}} \ket{\varphi_c \bar{\varphi}_c (\varphi\varphi^{*})^\textrm{S}} |^{2} ~,
\end{equation}
\begin{equation}
I_{\mathrm{T},\tilde{c}} = | \bra{\varphi_c \varphi \bar{\varphi}} a_{\tilde{c}} \ket{\varphi_c \bar{\varphi}_c (\varphi\varphi^{*})^\textrm{T}} |^{2} ~.
\end{equation}
\textcolor{black}{
The final shake-down state $\ket{\varphi_c \cdots\varphi \bar{\varphi}}$ is a doublet state,
and we assume in the following that $M_{S}=+\frac{1}{2}$ without loss of generality 
(the same reasoning applies to $M_{S}=-\frac{1}{2}$). 
Thus, only states (after the annihilation of one core electron) with 
the same
spin projection can have a non-zero overlap with the shake-down state. 
It is not possible to obtain such a doublet spin projection 
by coupling a $\beta$ core electron $\bar{\varphi}_{c}$ with the singlet-coupled valence electrons $(\varphi \varphi^*)^{\mathrm{S}}$. Consequently,
}
\begin{eqnarray}
\nonumber
    I_{\mathrm{S},c} = | \bra{\varphi_c \varphi \bar{\varphi}} a_{c} \ket{\varphi_c \bar{\varphi}_c (\varphi\varphi^{*})^\textrm{S}} |^{2} 
    \\\label{eq:s1}
    = | \bra{\varphi_c \varphi \bar{\varphi}} \bar{\varphi}_c (\varphi\varphi^{*})^\textrm{S} \rangle |^{2} = 0 ~.
\end{eqnarray} 
\textcolor{black}{On the other hand, we may construct this doublet by removing a $\beta$ core electron, in which case the relative intensity becomes}
\begin{align} 
\label{eq:s2}
    \begin{split}
    I_{\mathrm{S},\bar{c}} &= | \bra{\varphi_c \varphi \bar{\varphi}} a_{\bar{c}} \ket{\varphi_c \bar{\varphi}_c (\varphi\varphi^{*})^\textrm{S}} |^{2} 
        \\
        & = \tfrac{1}{2} | \bra{\varphi_c \varphi \bar{\varphi}} (\varphi_c \varphi\bar{\varphi}^{*} - \varphi_c \bar{\varphi} \varphi^{*}) \rangle |^{2} 
        \\
        &\approx \tfrac{1}{2} | \overlap{\varphi_c}{\varphi_c} \overlap{\varphi}{\varphi} \overlap{\bar{\varphi}}{\bar{\varphi}^{*}} + \overlap{\varphi_c}{\varphi_c} \overlap{\varphi}{\varphi^{*}} \overlap{\bar{\varphi}}{\bar{\varphi}} |^{2} \\
        &\approx \tfrac{1}{2} | \overlap{\bar{\varphi}}{\bar{\varphi}^{*}} +  \overlap{\varphi}{\varphi^{*}} |^{2} = |\overlap{\varphi}{\varphi^{*}}|^2
        \neq 0 ~,
    \end{split}
\end{align}
\textcolor{black}{
where we approximated the state overlaps to leading order with respect to the orbital overlap matrix elements, and used that $\overlap{\varphi_c}{\varphi_c}\sim 1$ and $\overlap{\varphi}{\varphi} \sim 1$.} 

\textcolor{black}{In the case of an initial triplet valence-excited state, an $\alpha$ and a $\beta$ core electron can only couple to the $M_{S}=0$ and $M_{S}=+1$ spin projections of the triplet state to produce a doublet. Hence,}
\begin{align} \label{eq:T1}
    \begin{split}
        I_{\mathrm{T},c} &= | \bra{\varphi_c \varphi \bar{\varphi}} a_{c} \ket{\varphi_c \bar{\varphi}_c (\varphi\varphi^{*})^\textrm{T}} |^{2} 
        \\
        &=  | \bra{\varphi_c \varphi \bar{\varphi}} \bar{\varphi}_c \varphi\varphi^{*} \rangle |^{2} 
        \\
        &\approx  | \overlap{\varphi_c}{\varphi^{*}} \overlap{\varphi}{\varphi} \overlap{\bar{\varphi}}{\bar{\varphi}_c} |^{2} \\
        &\approx  | \overlap{\varphi_c}{\varphi^{*}} \overlap{\bar{\varphi}}{\bar{\varphi}_c} |^{2} \simeq 0 ~,
    \end{split}
\end{align}
\textcolor{black}{
where the second equality follows from coupling a $\beta$ core electron with the ($M_{S}=+1$)-component of the triplet state. 
Since the overlap between a core and valence orbital is expected to be very small, the product of such two overlaps will be vanishingly small. 
Lastly, we have}
\begin{align} 
\label{eq:T2}
    \begin{split}
        I_{\mathrm{T},\bar{c}} &= | \bra{\varphi_c \varphi \bar{\varphi}} a_{\bar{c}} \ket{\varphi_c \bar{\varphi}_c (\varphi\varphi^{*})^\textrm{T}} |^{2} 
        \\
        &= \tfrac{1}{2} | \bra{\varphi_c \varphi \bar{\varphi}} (\varphi_c \varphi \bar{\varphi}^{*} + \varphi_{c} \bar{\varphi} \varphi^* ) \rangle |^{2} 
        \\
        &\approx 
        \tfrac{1}{2} | \overlap{\varphi_c}{\varphi_c} \overlap{\varphi}{\varphi} \overlap{\bar{\varphi}}{\bar{\varphi}^{*}} - \overlap{\varphi_c}{\varphi_{c}} \overlap{\varphi}{\varphi^*} \overlap{\bar{\varphi}}{\bar{\varphi}} |^{2} \\
        &\approx 
        \tfrac{1}{2} | \overlap{\bar{\varphi}}{\bar{\varphi}^{*}} - \overlap{\varphi}{\varphi^*} |^{2} = 0 ~,
    \end{split}
\end{align}
\textcolor{black}{where the second equality follows from coupling an $\alpha$ core electron with the ($M_{S}=0$)-component of the triplet state, and the last equality is a consequence of the two contributions cancelling out.}

\textcolor{black}{To summarize, eqn.~\eqref{eq:s1}--\eqref{eq:T2} show that only molecules in a singlet valence-excited state at the core ionization event will give rise to a non-vanishing intensity. 
We speak of a propensity rule rather than a strict selection rule because we only consider leading orders of the squared Dyson amplitudes. 
Moreover, we note that configuration interaction 
and spin-orbit coupling may complicate the picture and somewhat weaken our rule.}

\begin{acknowledgement}
\textcolor{black}{We are grateful to the whole FERMI team for the dedicated support during the beamtime (proposal n. 20214068-Di Fraia), and we thank A. Gessini, M. Zamolo and A. Hervat for their outstanding technical support.}
S.C. and J.P. acknowledge support from the
Novo Nordisk Foundation Data Science Research Infrastructure 2022 Grant: A high-performance computing infrastructure for data-driven research on sustainable energy materials,
Grant no. NNF22OC0078009.
J.P. acknowledges financial support from the Technical University of Denmark within the Ph.D. Alliance Programme.
\textcolor{black}{S.C. thanks Hamburg's Cluster of Excellence ``CUI: Advanced Imaging of Matter'' for the
2024 Mildred Dresselhaus Prize}.
R.S.M. would like to thank the EPSRC (EP/X027635/1) and Leverhulme trust (RPG-2021-257) for financial support. H.J.T. thanks the UK XFEL hub for physical sciences and the University of Southampton for a PhD studentship.
R.Fo. and F.A. gratefully acknowledge support from the Linac Coherent Light Source, SLAC National Accelerator Laboratory, which is supported by the US Department of Energy, Office of Science, Office of Basic Energy Sciences, under contract no. DE-AC02-76SF00515.
D.M.P.H. is grateful to the Science and Technology
Facilities Council (United Kingdom) for financial support. D.R. acknowledges support from the National Science Foundation Grant no. PHYS-2409365
R.F. acknowledges financial support from the Swedish Research Council (grant number 2023-03464) and the Knut and Alice Wallenberg Foundation (grant numbers 2017.0104 and 2024.0120).



\end{acknowledgement}

\begin{suppinfo}

Supplementary information with further details on the experiment and calculations is available:

\begin{itemize}
  \item Time-resolved core experimental analysis details: spectrometer energy calibration, delay calibration, TR-XPS, Bootstrapped analysis, Fit Equations \& Time constants. 
  \item Time-resolved valence experimental analysis details: binding energy calibration, TR-UPS, Kinetic Fits. 
  \item Computational information and results: Geometries, Active spaces, Computed XPS spectra, Assignment of spectroscopic features, Additional CCSD results.
\end{itemize}

\end{suppinfo}

\bibliography{achemso-demo}

\begin{mcitethebibliography}{41}
\bibitem[Not()]{Note-1}
Other processes such as shake-off are possible but we limit the discussion here for the sake of brevity.\relax
\mciteBstWouldAddEndPunctfalse
\mciteSetBstMidEndSepPunct{\mcitedefaultmidpunct}
{}{\mcitedefaultseppunct}\relax
\EndOfBibitem
\end{mcitethebibliography}

\end{document}


\maketitle
\clearpage
\tableofcontents
\clearpage

\section{Time-resolved core experimental and data analysis details}\label{S1}

\begin{figure}[hbt]
    \centering
    \includegraphics[width=\linewidth]{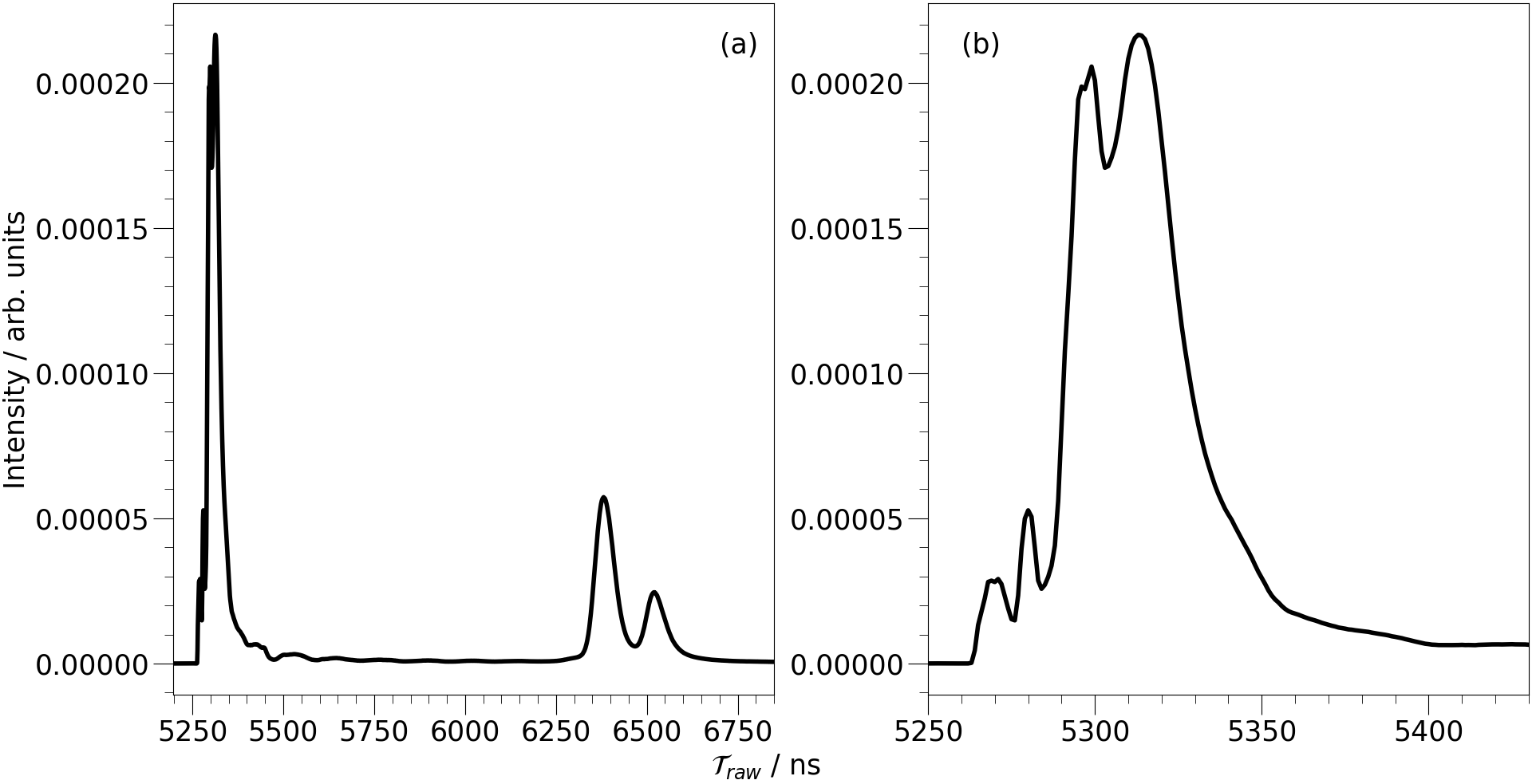}
    \caption{a) Time-of-flight photoelectron spectra of CS$_2$ probed using a 179.948~eV photon energy and a retardation voltage of 4~V. The S 2p$_{3/2}$ and S 2p$_{1/2}$ peaks that are the main focus of the work presented are located at TOF of 6380 and 6520~ns, respectively. The peaks between 5200 and 5360~ns relate to higher energy electrons associated with ionisation of the helium carrier gas, valence ionisation of \ce{CS2} and secondary Auger-Meitner emission of \ce{CS2} following core ionisation. An expanded view of the high energy region is plotted in b).}
    \label{fig:XPS TOF}
\end{figure}

Raw data were recorded on a shot-to-shot basis at the FEL repetition rate, 50 Hz, while the Seed Laser for Users (SLU) operated at a repetition rate of 25 Hz. As such, the structure of the data alternates sequentially between SLU-on and SLU-off. Raw time-of-flight (TOF) traces consist of the negative-going output signal of the magnetic bottle spectrometer, acquired by a CAEN VX1751 digitizer with 10-bit resolution, 1 V dynamic range, 1 ns sampling time. The sign of the digital traces is then inverted and a threshold of 3 mV above baseline is applied to each trace to reduce electrical noise. At every delay, the SLU-on and SLU-off spectra for each included shot are summed to produce separate delay-dependent SLU-on and SLU-off spectra. The SLU-on and SLU-off spectra obtained at each delay are subsequently normalised by the aggregate FEL pulse energy of the included shots. This process is followed for both the S 2p TR-XPS (Fig.~\ref{fig:XPS TOF}) and the valence data sets. Note that the digitizer is triggered in advance of the FEL pulse, and the time of flight $\mathcal{T}$ is calculated as $\mathcal{T}_\mathrm{raw} - \mathcal{T}_0$, where $\mathcal{T}_0 \approx 5000$~ns is the arrival time of the FEL pulse; the spectra in Fig.~\ref{fig:XPS TOF} are plotted versus $\mathcal{T}_\mathrm{raw}$.

\subsection{Spectrometer energy calibration}\label{SI1.1}

The TOF axis is calibrated using the known peaks associated with ionisation of the ground state \ce{CS2} molecules and the helium carrier gas. Measurements at an X-ray photon energy of 179.948~eV (6.89~nm) and spectrometer retardation voltages of 4~V and 100~V were used and a summary of the peaks used is given in Table~\ref{tab:XPS calibration}. The kinetic energy values for the Auger-Meitner transitions were taken directly from \cite{Hedin2009} and the retardation voltages were taken into account. Kinetic energy values for all other features were calculated using reported binding energies taking into account the photon energy and respective retardation voltage. These reference binding energy values for peaks originating from \ce{CS2} were taken from \cite{H_Wang_2001}, and \cite{Baltzer96}, and for helium from \cite{PhysRevLett.105.063001}. 

The following equation is used to convert electron time-of-flight to kinetic energy:
\begin{table}[b]
\begin{tabular}{|l|l|l|l|l|l|}
\hline
Species & Feature            & $\mathcal{T}_\mathrm{raw}$ / ns & $E_\mathrm{k}$ / eV & $V_\mathrm{ret}$ / V & Reference Features\\ \hline
CS$_{2}$ & S 2p$_{1/2}$      & 6520     & 4.848      & 4    & S$({}^2{\mathrm{P}}_{1/2})$~\cite{H_Wang_2001}\\ \hline

CS$_{2}$ & S 2p$_{3/2}$       & 6380     & 5.948     & 4    & S$({}^2{\mathrm{P}}_{3/2})$~\cite{H_Wang_2001} \\ \hline
CS$_{2}$ & S L$_{2,3}$VV Auger peak  & 5521 & 42.1& 100                     &      $\mathrm{(S~2p)^{-1}~{}^2P_{3/2}\rightarrow(1\pi_g)^{-2};a {}^1\!\Delta_{g}}$ \cite{Hedin2009}\\ \hline
He & 1s ${}^2{\mathrm{S}}_{1/2}$          & 5458     & 55.361                 & 100                       & He~1s $ {}^2\mathrm{S}_{1/2}$~\cite{PhysRevLett.105.063001}\\ \hline
CS$_{2}$ & Valence onset& 5406  & 69.868                 & 100                       & $\mathrm{2 \pi_g^{-1} ~(X~{}^2\Pi_g)}$ ~\cite{Baltzer96}\\ \hline
CS$_{2}$ & S L$_{2,3}$VV Auger   peak  & 5295 
& 138.1 & 4                         & $\mathrm{(S~2p)^{-1}~{}^2\mathrm{P}_{3/2}
\rightarrow(1\pi_g)^{-2};a {}^1\!\Delta_g}$  \cite{Hedin2009}                   \\ \hline
He & 1s ${}^2\mathrm{S}_{1/2}$             & 5280     & 151.361                & 4                         & He~1s ${}^2\mathrm{S}_{1/2}$~ \cite{PhysRevLett.105.063001}\\ \hline
CS$_{2}$ & Valence onset  
& 5268     & 165.868  & 4        & 
$\mathrm{2 \pi_g^{-1}~(X~^2\Pi_g)}$  ~\cite{Baltzer96}\\ \hline
\end{tabular}
\caption{Data used in the energy calibration of the magnetic bottle spectrometer.
    \label{tab:XPS calibration}}
\end{table}
\newpage
\begin{align} 
E_{k} =  &\frac{1}{2} m_\mathrm{e}\left( \frac{L}
{\mathcal{T}} \right)^2 \nonumber \\
= &\left( \frac{1686.065\ L/\mathrm{m}}
{\mathcal{T/\mathrm{ns}}} \right)^2 \mathrm{eV}
\label{eq:XPS eKE calibration}
\end{align}
where $m_\mathrm{e}$ is the mass of an electron, $L$ is the flight length, and $\mathcal{T} = \mathcal{T}_\mathrm{raw} - \mathcal{T}_0$ is the electron time-of-flight, as defined above.  Upon calibration of the equation to the data within Table~\ref{tab:XPS calibration}, we obtain the best-fit values $L = 1.97 \pm 0.02$~m, $\mathcal{T}_0 =  5010 \pm 3.07$~ns.

\begin{figure}[hbt]
    \centering
    \includegraphics[width=0.5\linewidth]{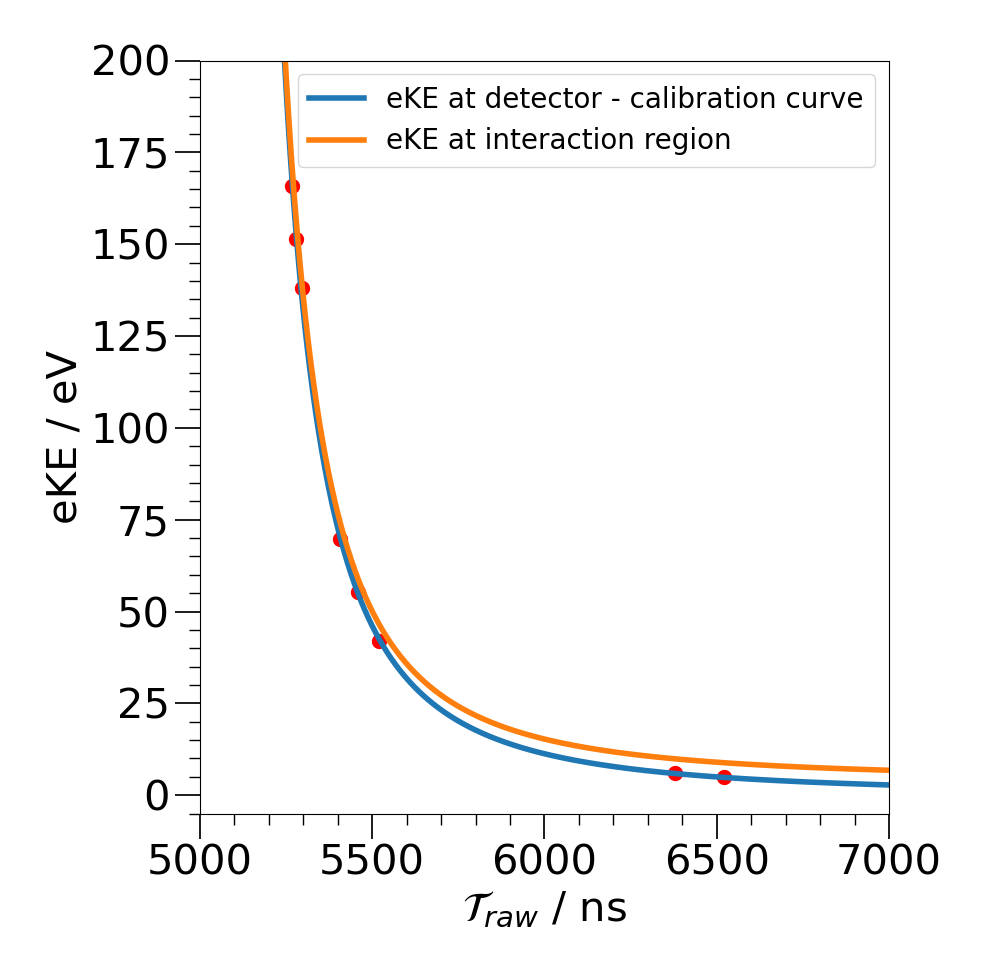}
    \caption{Time-of-flight to electron kinetic energy calibration curve resulting from the calibration points in Table~\ref{tab:XPS calibration}. The red data points are the calibration points defined by the data in Table~\ref{tab:XPS calibration}. The blue curve represents the fit of the measured kinetic energy at the detector position. The orange curve has the 4~V retardation voltage added to the energy and gives the kinetic energy of the electron upon ionisation, as displayed in the manuscript.}
    \label{fig:XPS eKE calibration}
\end{figure}

The resulting energy calibration curve is presented in Fig.~\ref{fig:XPS eKE calibration}, plotted versus $\mathcal{T}_\mathrm{raw}$. Two curves are presented, giving the electron kinetic energy $E_{\mathrm{k}}$ as detected, taking into account the retardation applied, $V_\mathrm{ret}$, and a second curve that defines the energy of the electrons $E_\mathrm{k,0} = E_\mathrm{k} + |e| V_\mathrm{ret}$ at the point of ionisation ($e<0$ is the value of the electron charge). The values presented in the manuscript are the kinetic energies of the electron upon ionisation, $E_\mathrm{k,0}$, prior to the effect of the retardation voltage. 

Following conversion to kinetic energy, the photoelectron-time-of-flight distribution is scaled by the Jacobian factor $|\mathrm{d}E_\mathrm{k}/\mathrm{d}\mathcal{T}| \propto E_\mathrm{k}^{-3/2}$ to obtain the correct photoelectron-energy distribution.

\subsection{Delay calibration}\label{SI1.2}
\begin{figure}[hbt]
    \centering
    \includegraphics[width=0.75\linewidth]{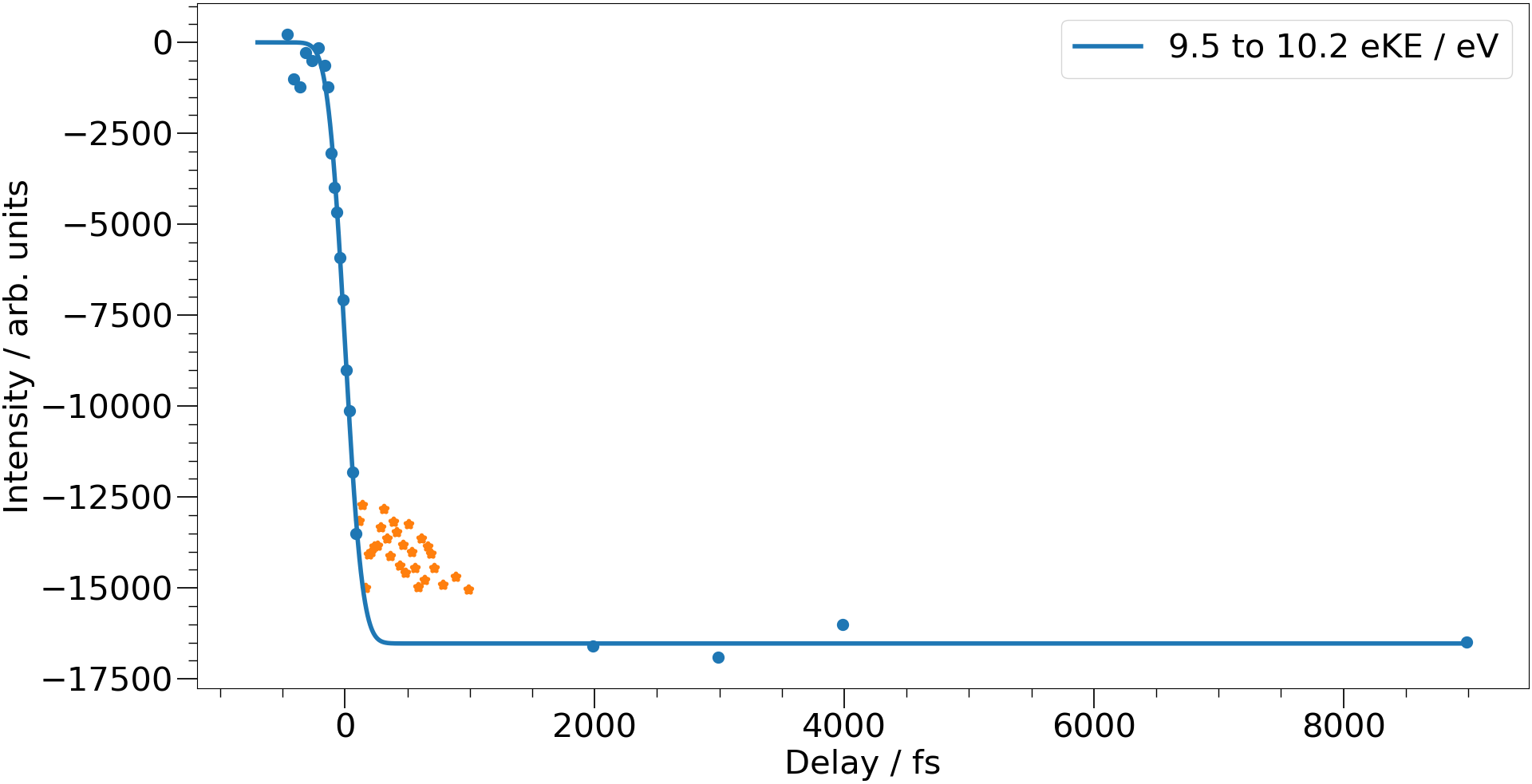}
    \caption{Time-dependent integrated intensity of the X-ray photoelectron spectrum of \ce{CS2} between 9.5-10.2~eV. The blue and orange data points define the delay points that are included or excluded from the error function fit respectively.}
    \label{fig:XPS GSD}
\end{figure}

The time-zero for the pump-probe delay is experimentally defined by depletion of the ground state XPS signal located at an eKE between 9.5~eV and 10.2~eV. The integrated intensity over the $9.5-10.2$~eV range is plotted in Fig.~\ref{fig:XPS GSD} alongside a fit to an error function depletion. Due to overlapping contributions from excited state signals, the delays between 110 and 1900 fs are removed from the fit and are plotted in orange in Fig.~\ref{fig:XPS GSD}. The fit to the remaining points provides a time-zero that is used in all plots and has an uncertainty of $\pm$~4.1~fs. The fit also provides a pump-probe cross-correlation width ($\sigma$) of 108 $\pm$ 5.8~fs, consistent with the quadrature sum of the individual pulse durations. We use this value in the later kinetic fits.

\subsection{TR-XPS}\label{SI1.3}

\begin{figure}[hbt]
    \centering
    \includegraphics[width=\linewidth]{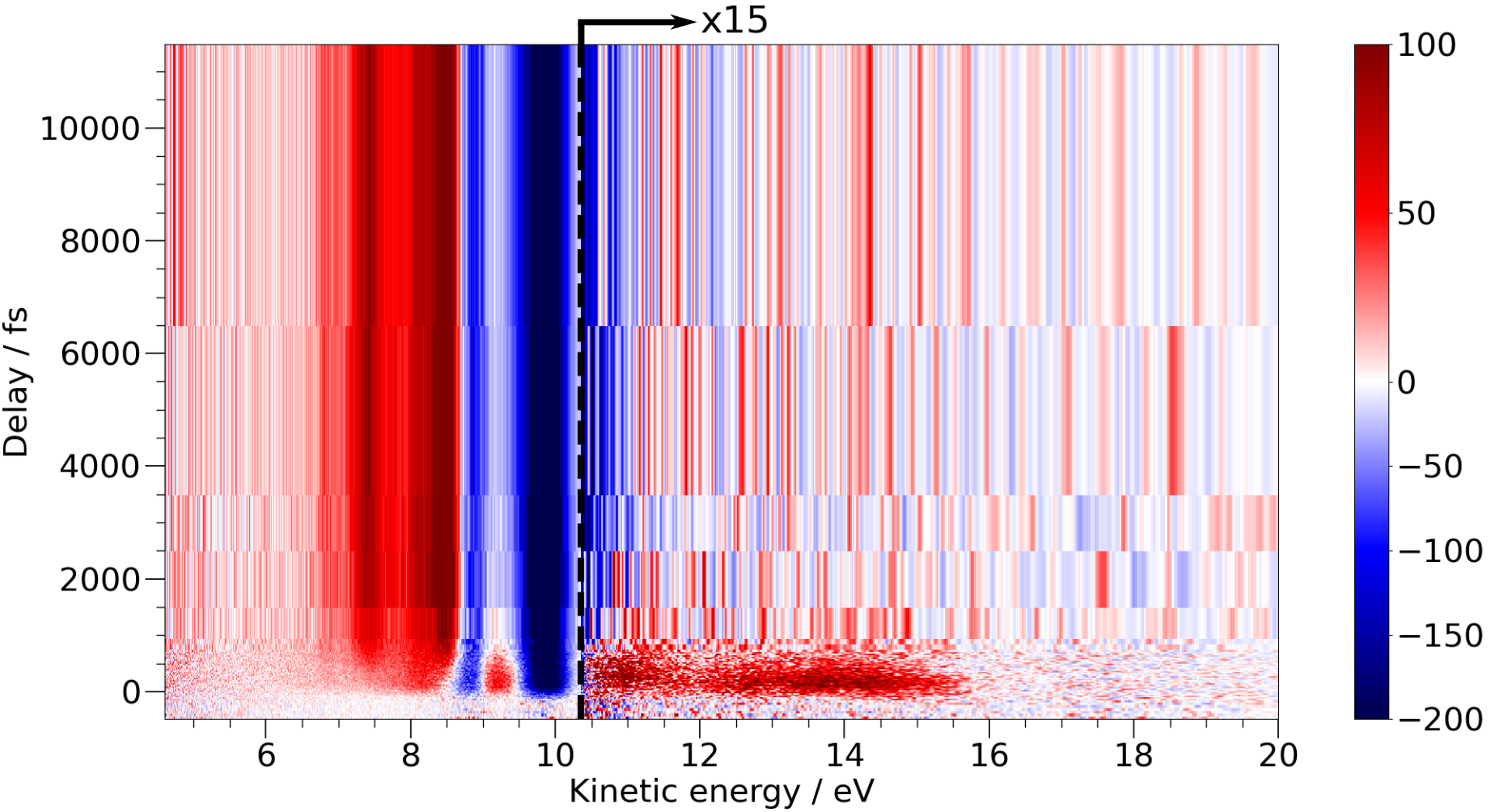}
    \caption{Time-resolved differential X-ray photoelectron spectrum of \ce{CS2} obtained following 200 nm excitation and ionisation with a 179.9 eV probe.}
    \label{fig:XPS full contour}
\end{figure}

In the manuscript we present the time-resolved data up to 2.3~ps. The full measurement covered a time range up to 9~ps. For completeness, in Fig.~\ref{fig:XPS full contour} we present the experimental TR-XPS data, over the full delay range measured, $-0.46$ to $+9$~ps. A Gaussian filter has been applied to the data using the \texttt{scipy.ndimage.gaussian\_filter} function with a standard deviation of 1 (data point, 1 ns in TOF) for the Gaussian kernel applied along the energy axis. This process is applied to all experimental spectra and contour maps presented in this manuscript.

\subsection{Bootstrapped analysis}\label{SI1.4}

Confidence regions for the integrated intensities of the TR-XPS presented in Fig.~3 of the main manuscript were estimated via bootstrapping. By resampling  the data on a shot-by-shot basis with replacement, allowing for repeats and omissions, a series of 5000 integrated intensities were obtained at each delay. From the distributions the standard deviations were obtained for each delay, with the 2$\sigma$ range presented in Fig.~3 of the main paper.

\subsection{Fit Equations \& Time constants}\label{SI1.5}
The integrated intensities plotted in Fig.~3 of the main paper are fit to various functions. Signals associated with changing populations of the ground and excited states of \ce{CS2} are fit to a simple kinetic model that describes the flow of population between the ground and two separate electronically excited states. The principles of the model are outlined in panel (e) of Fig.~3 in the main manuscript. Interaction of the pump pulse with the molecule, transfers some population from the ground state, $\ket{0}$, to an electronically excited excited state, $\ket{1}$. State $\ket{1}$ exponentially decays with a rate constant, $k_{1}$, into a second electronically excited state, $\ket{2}$. Population in state $\ket{2}$ then exponentially decays with a rate constant of $k_{2}$. Assuming first order kinetics, equations that model the changes in state populations, $I_{0}$, $I_{1}$ and $I_{2}$, and therefore reflect changes in the signal intensities associated with these populations are given in equations \ref{eq:I_{0}} - \ref{eq:I_{2}} respectively.


\begin{equation} \label{eq:I_{0}}
    I_{0} = -A_0 \left( 1+\mathrm{erf}\frac{\Delta t}{\sqrt2 \sigma} \right)  
\end{equation}\\
\begin{equation} \label{eq:I_{1}}
\begin{split}
    I_{1} = A_1  e^{-k_{1}\Delta t} 
    e^{ \frac{ \left(\sigma k_{1} \right)^2}{2} }
    \left[ 1+\mathrm{erf} \left(\frac{\Delta t - 
    \sigma^2 k_{1}}{\sqrt2 \sigma} \right) \right]
\end{split}
\end{equation}\\

\begin{equation} \label{eq:I_{2}}
\begin{split}
    I_{2} = A_2 \left\{e^{-k_{2 }\Delta \textit{t}} 
    e^{ \frac{ \left(\sigma k_{2 } \right)^2}{2} }
    \left[ 1+\mathrm{erf} \left(\frac{\Delta \textit{t} - 
    \sigma^2 k_{2}}{\sqrt2 \sigma} \right) \right]- e^{-k_{1}\Delta \textit{t}} 
    e^{ \frac{ \left(\sigma k_{1} \right)^2}{2} }
    \left[1 +\mathrm{erf} \left(\frac{\Delta \textit{t} - 
    \sigma^2 k_{1}}{\sqrt2 \sigma} \right) \right] \right\}
\end{split}
\end{equation}
where the $A_n$ terms define relative amplitudes, $\Delta t$ is the pump-probe delay, and $\sigma$ defines the laser cross-correlation. $\sigma$ has a fixed value of 108~fs in the fits as defined by the time-zero fit outlined in Section~\ref{SI1.2}. The results of the fits are summarised in Table~\ref{tab:XPS kinetic fit constants} and plotted in Fig.~3 of the main manuscript. 

\begin{table}[]
\centering
\begin{tabular}{|l|l|l|l|}
\hline
 Energy Range / eV & Fit Function & $(1/k_{1})$ / fs & $(1/k_{2})$ / fs\\ \hline
  13.2 – 15.7 & $I_1$ & 345.3~$\pm$~11.6& N/a \\ \hline
10.5 – 12  & $I_0 + I_2$ & 345.3 (Fixed)& 166.8~$\pm$~15.4\\ \hline
9 – 9.5  & $I_0 + I_2$  & 20.8~$\pm$~0.4& 739.1~$\pm$~34.2\\ \hline
\end{tabular}
\caption{Parameters obtained from fits to the measured shake-down and excited state XPS signals plotted in Fig.~3 (a) – (c) of the main paper.}
    \label{tab:XPS kinetic fit constants}
\end{table}

The intensity of the CS product signal, $I_\mathrm{CS}$ is separately fit to a sum of logistic functions of the form given in equation \ref{eq:I_{ logistic}}.


\begin{equation} \label{eq:I_{ logistic}}
\begin{split}
    I_\mathrm{CS} =\sum_n A_n \left[\frac{1}{1+ e^{-k_n \left( \Delta t - t_{n}^{\mathrm{offset}} \right)} }  \right ]
\end{split}
\end{equation}
where terms have the same meaning as above, $k_n$ is the exponential time-constant, and $t_{n}^\mathrm{offset}$ is the offset delay from time zero. The sum is restricted to two terms in the XPS fit with the parameters extracted from the fit given in Table~\ref{tab:XPS logistic fit constants}.

\begin{table}[]
\centering
\begin{tabular}{|l|l|l|l|l|}
\hline
Energy range / eV & $t_1^{\mathrm{offset}}$ / fs & $(1/k_1$) / fs & $t_2^{\mathrm{offset}}$ / fs & $(1/k_2)$ / fs \\ \hline
$7.2-7.6$ & $19.7 \pm 10.0$& $54.9\pm 16.7$& $685.5 \pm 23.2$& $245.6 \pm 19.5$\\ \hline
\end{tabular}
\caption{Parameters obtained from the double logistic function fit to the \ce{CS} fragment signal between $7.2 – 7.6$~eV. The fit is plotted along with the data in Fig.~3 (d) of the main paper.}
    \label{tab:XPS logistic fit constants}
\end{table}

\section{Time-resolved valence experimental and data analysis details}\label{SI2}


Time-resolved valence photoelectron spectroscopy (TR-UPS) data has also been recorded for comparison to previously published measurements of the valence photoelectron spectrum and to show consistency with the XPS measurements. As described in the experimental method section of the main manuscript, the FERMI FEL-2 radiation is produced in a double stage cascade configuration. To record valence data at higher resolution we ``switch off'' the second stage, obtaining radiation at 20.67 nm (59.98 eV) with the first stage alone. Importantly the pump conditions were identical to those used in the XPS measurements reported in the main manuscript.

\subsection{Binding energy calibration}\label{SI2.1}

\begin{table}[]

\begin{tabular}{|l|l|l|l|}
\hline
Feature & $\mathcal{T}_\mathrm{raw}$ / ns & $V_\mathrm{ret}$ / V & Reference features\\ \hline
X       & 5883     & 35                        & $\mathrm{2\pi_g^{-1} ~(X~ ^2\Pi_g)}$  ~\cite{Baltzer96}\\ \hline
A       & 5980     & 35                        & $\mathrm{2\pi_u^{-1} ~(A~ ^2\Pi_u)}$   ~\cite{Baltzer96}\\ \hline
B       & 6037     & 35                        & $\mathrm{5\sigma_u^{-1} ~(B~ ^2\Sigma_u^+)}$    ~\cite{Baltzer96}\\ \hline
C       & 6127     & 35                        & $\mathrm{6\sigma_g^{-1} ~(C~ ^2\Sigma_g^+)}$   ~\cite{Baltzer96}\\ \hline
Helium  & 8640     & 35                        & He~1s $ {}^2\mathrm{S}_{1/2}$~ \cite{PhysRevLett.105.063001}\\ \hline
\end{tabular}
\caption{Time-of-flight to binding energy calibration points used in the calibration of Fig.~\ref{fig:Valence eKE calibration}.}
    \label{tab:Val calibration}
\end{table}

The valence photoelectron spectroscopy measurements are plotted against binding energy to aid comparison with the existing literature. The calibration of the spectrometer followed the same procedure as outlined for the XPS measurements, but used the known valence ionisation peaks outlined in Table~\ref{tab:Val calibration}, with literature values taken from~\cite{Baltzer96} and ~\cite{PhysRevLett.105.063001}.  A retardation voltage (35 V, nominal) was applied to improve spectral resolution. Based on Einstein's photoelectric equation, and on Eq.~\ref{eq:XPS eKE calibration}, we define the photoelectron binding energy as: 
\begin{equation} \label{eq:UPS eBE calibration}
    E_\mathrm{b} = hv - \frac{1}{2} m_\mathrm{e} \left( \frac{L}{\mathcal{T}} \right)^2 - |e|{V_\mathrm{ret}}
\end{equation}\\  

\begin{figure}[hbt]
    \centering
    \includegraphics[width=0.5\linewidth]{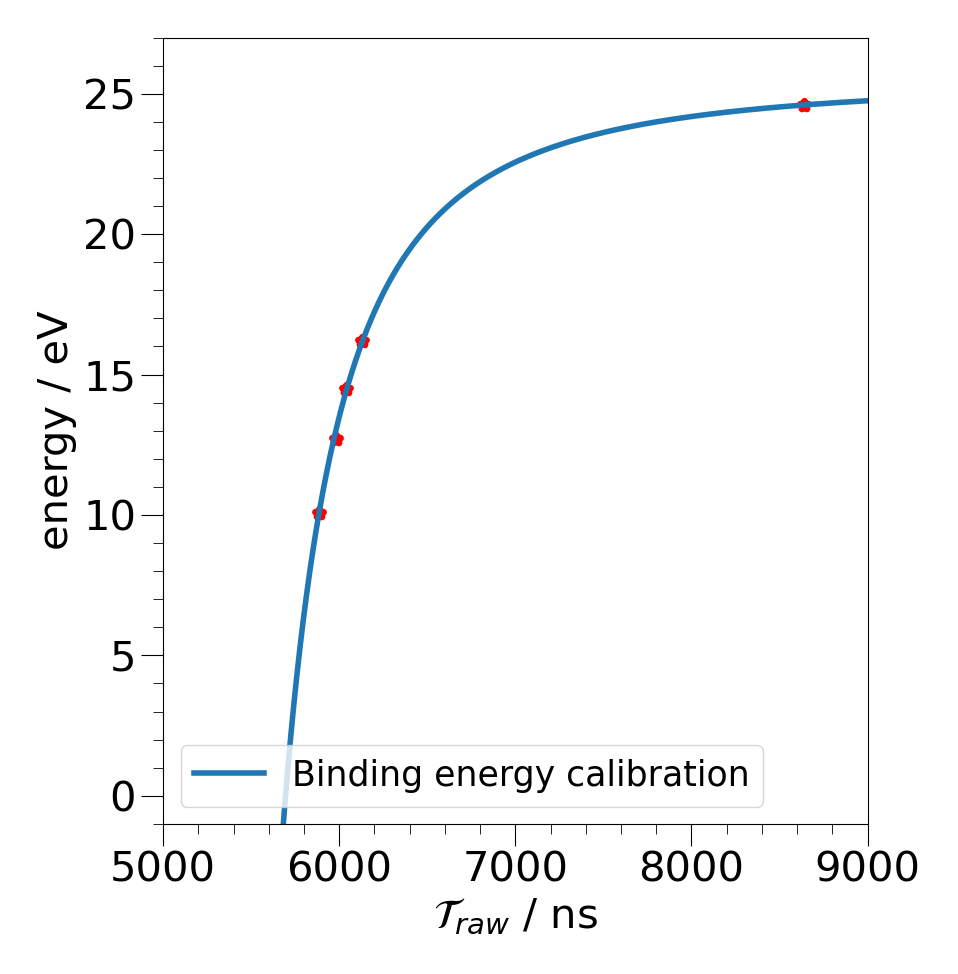}
    \caption{Time-of-flight to binding energy calibration curve (blue line) resulting from the calibration points in Table~\ref{tab:Val calibration} (red dots).}
    \label{fig:Valence eKE calibration}
\end{figure}

Upon fitting equation~\ref{eq:UPS eBE calibration} to the data in table~\ref{tab:Val calibration}, we obtain best fit values to be: $L = 1.99 \pm 0.14$~m, $\mathcal{T}_0 = 5030 \pm 60.54$~ns, $V_\mathrm{ret} =  34.52 \pm 0.29$~eV, with the FEL photon energy $h\nu$ fixed at the nominal value, 59.98 eV. The resulting TOF to binding energy conversion is plotted in Fig.~\ref{fig:Valence eKE calibration}.

\subsection{TR-UPS}\label{SI2.2}


\begin{figure}[hbt]
    \centering
    \includegraphics[width=\linewidth]{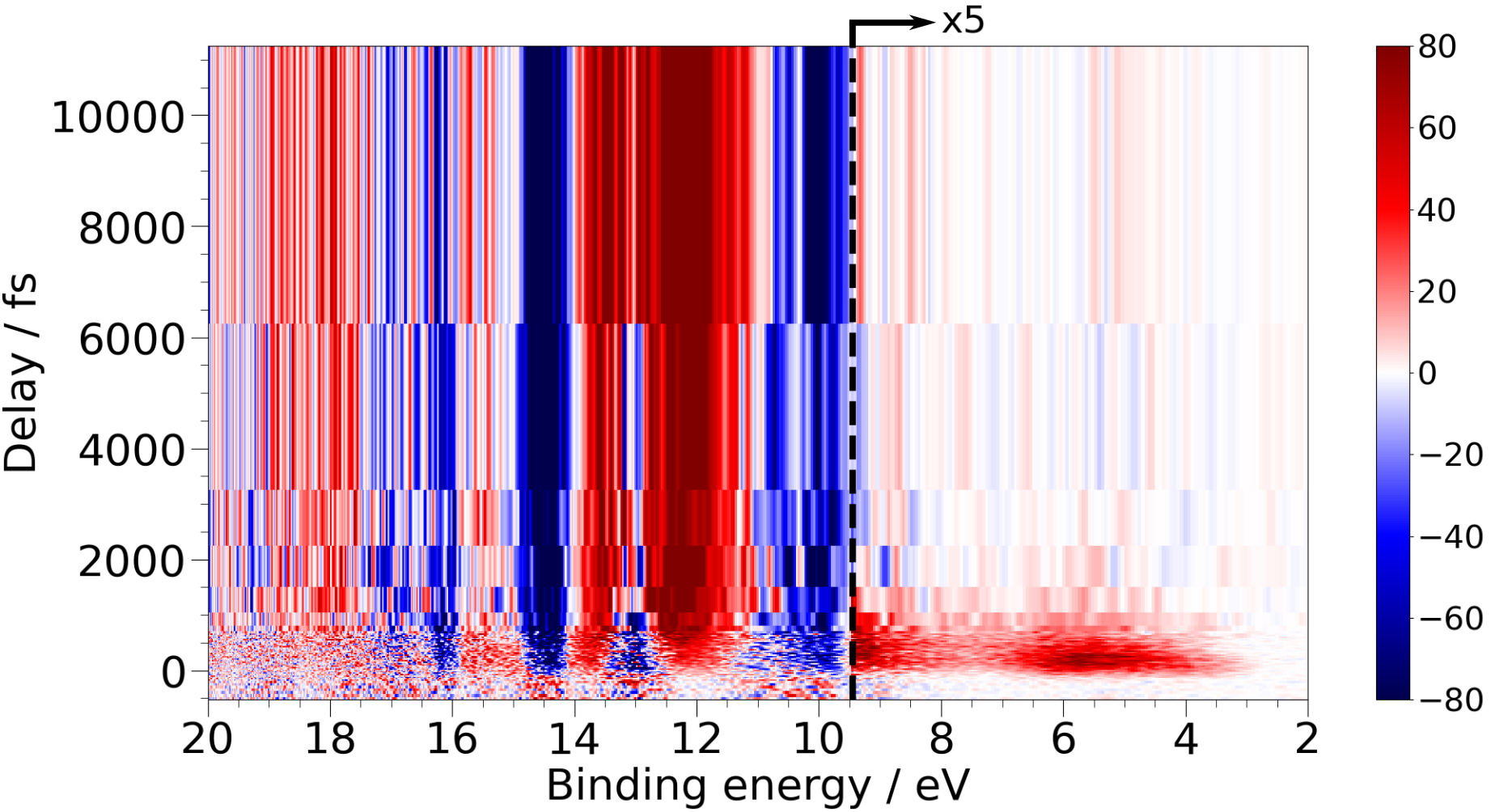}
    \caption{Time-resolved differential UV photoelectron spectrum of CS$_2$ obtained following 200 nm excitation and ionisation with a 60~eV probe up to 9 ps.}
    \label{fig:UPS}
\end{figure}

In Fig.~\ref{fig:UPS} we present the differential, pump-on minus pump-off, valence photoelectron spectra obtained at pump-probe delays out to 9~ps. Blue regions in the spectra relate to depletions of the ground state signal, while the red regions are enhancements due to excited state populations (between 3.2-9.5~eV) or product state formation (between 11.4-14.1~eV). The time-zero point for the pump-probe delay is defined by the kinetic fit (eq~\ref{eq:I_{1}}) to the signal associated with population of the initial excited state as measured at electron binding energies between 3.2 – 7.5~eV.


\subsection{Kinetic fits}\label{SI2.3}

As with the analysis of the XPS data we extract time-dependent intensity profiles for various energy regions to explore the time-scales and kinetics of the changes observed. 

\begin{figure}[hbt]
    \centering
    \includegraphics[width=0.5\linewidth]{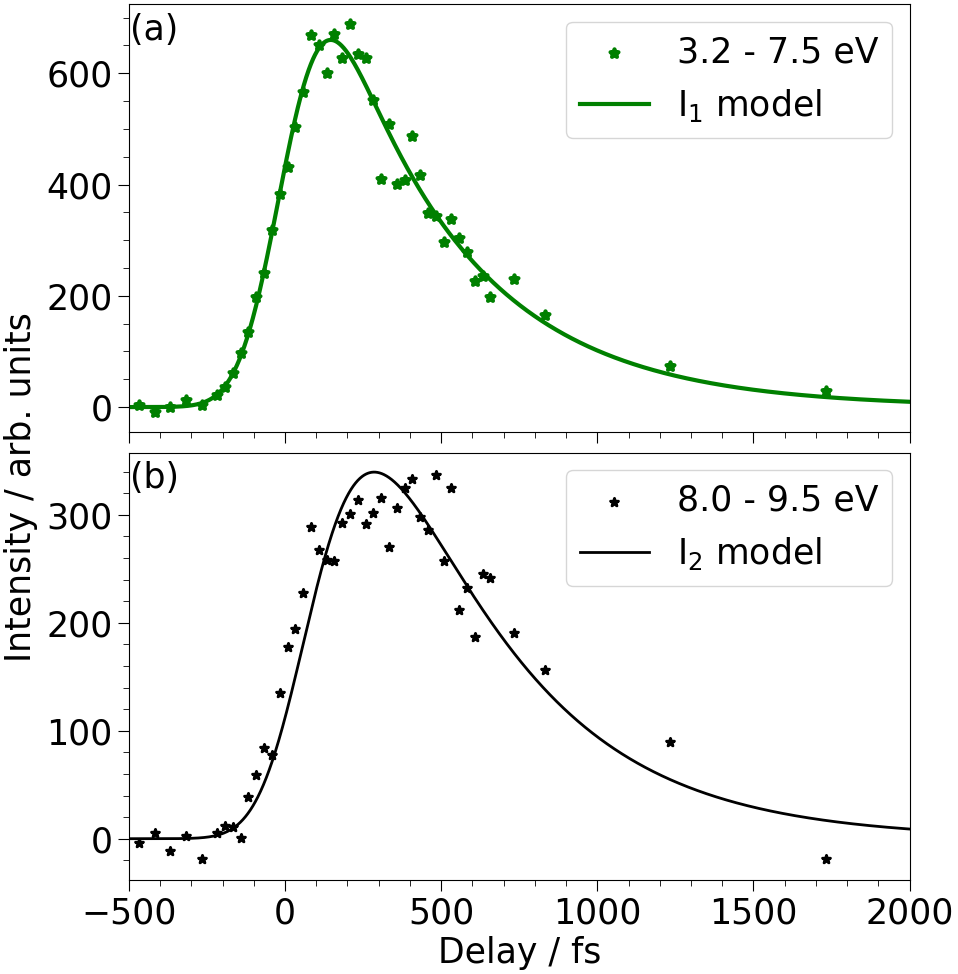}
    \caption{Integrated photoelectron intensity of the Valence photoelectron bands observed at 3.2 – 7.5~eV (a) and 8.0 – 9.5~eV (b) (Binding energy). Each intensity profile is fit to the equations described in the above SI text.}
    \label{fig:UPSLine}
\end{figure}

In line with previous work~\cite{Smith2018,Karashima2021}, we take the signals observed at binding energies below 9.5~eV as indicative of electronically excited states. Like the shake-down signals seen in the XPS measurements, the excited state signal in the valence measurement is split into two spectrally and temporally separate regions. The intensity profiles associated with the 3.2 – 7.5~eV and 8.0 - 9.5~eV binding energy regions are plotted in panels (a) and (b) of Fig.~\ref{fig:UPSLine} respectively. To directly compare the time-dependence of the valence signals with those in the shake down region, we fit the valence data to the same kinetic equations outlined in Section~\ref{SI1.5}. The lower binding energy region, 3.2 – 7.5~eV, is fit to equation $I_{1}$, representing the initial excited state population, Fig.~\ref{fig:UPSLine}(a), while the higher binding energy region, 8.0 - 9.5~eV, is fit to equation $I_{2}$ as a secondary populated region, Fig.~\ref{fig:UPSLine}(b). The extracted fit parameters are reported in Table~\ref{tab:UPS time constants}. The extracted time constants, 1/$k_{1}$ and 1/$k_{2}$, were 422.2~$\pm$~26.5~fs and 143.6~$\pm$~14.1~fs respectively, which closely match the values obtained from the shake-down regions indicating they derive from the same excited state populations. 

\begin{table}[]
\centering
\begin{tabular}{|l|l|l|l|}
\hline
$E_\mathrm{b}$ range / eV   & $\sigma$ / fs  & $(1/k_1)$ / fs & $(1/k_2)$ / fs\\ \hline
$3.2-7.5$   & $109.6\pm 6.2$  & $422.2\pm26.5$ & N/a\\ \hline
$8.0-9.5$   & 109.6, fixed & 422.2, fixed & $143.6\pm14.1$ \\ \hline
\end{tabular}

\caption{Parameters obtained from fits to the measured excited state UPS signals plotted in Fig.~\ref{fig:UPSLine}}
    \label{tab:UPS time constants}
\end{table}

Time constants for all of the valence fits are presented in Table~\ref{tab:UPS time constants}.
These values match those previously obtained in valence photoelectron spectroscopy measurements of \ce{CS2}~\cite{Smith2018,Karashima2021}, confirming that the general experimental approach is consistent with previous works. 

\newpage

\section{Computational information and results} \label{SI3}

\subsection{Geometries} \label{SI3.1}

The ground-state equilibrium structure of CS$_2$ was taken from Ref. \citenum{brown1999excited}. The excited-state geometries of CS$_2$ have been optimized at the (frozen-core) EOM-CCSD/aug-cc-pVDZ level of theory. The structure of CS$_{2}^{+}$ was optimized using (frozen-core) EOM-IP-CCSD/aug-cc-pVDZ. All geometry optimizations have been performed with Q-Chem (version 6.1)~\cite{Qchem541} with default convergence criteria and enforcing $C_{2\mathrm{v}}$ point group symmetry. The ground-state structure of CS was taken from Ref.~\citenum{huber1979molecular}. The bond lengths and bond angles of CS$_2$, CS$_{2}^{+}$, and CS are tabulated in Tab.~\ref{tab:geom}.

\vspace{0.5cm}

\begin{table}[htp!]
\captionsetup{width=10cm}
    \centering
    \normalsize
    \begin{tabular}{l c c c c}
         \hline
         \hline
         Molecule & State & $R_{\mathrm{CS}}$ [Å] & $\theta_{\mathrm{SCS}}$ [deg.] & Taken from \\ 
         \hline 
         CS$_{2}$ & GS & 1.555000 & 180.000000 & Ref.~\citenum{brown1999excited}\\
          & $1~{}^{1}\!A_{2}$ & 1.651757 & 137.092077 & This work\\
          & $2~{}^{1}\!A_{2}$ & 1.633878 & 176.234559 & This work\\
          & $1~{}^{1}\!B_{2}$ & 1.653275 & 129.426236 & This work\\
          & $2~{}^{1}\!B_{2}$ & 1.671133 & 150.501754 & This work\\
          & $1~{}^{3}\!A_{2}$ & 1.653716 & 136.278039 & This work\\
          & $2~{}^{3}\!A_{2}$ & 1.632599 & 166.253898 & This work\\
          & $1~{}^{3}\!B_{2}$ & 1.650480 & 124.773319 & This work\\
          & $2~{}^{3}\!B_{2}$ & 1.627617 & 168.543120 & This work\\
          \hline
         CS$_{2}^{+}$ & GS & 1.575570 & 180.000000 & This work\\
         \hline
         CS & GS & 1.534942 & 180.000000 & Ref.~\citenum{huber1979molecular}\\
         \hline
         \hline
    \end{tabular}
    \caption{Bond lengths and bond angles of CS$_2$, CS$_{2}^{+}$, and CS.}
    \label{tab:geom}
\end{table}

\clearpage

\subsection{Active Spaces}\label{SI3.2}

The Hartree-Fock electronic configuration of CS$_2$ at linear and bent geometry is summarized in Table~\ref{tab:GS-HFconfig}.

\begin{table}[htpb!]
    \centering
    \begin{tabular}{c|c|c|c}
    \hline\hline
        $D_{\infty \mathrm{h}}$ & $D_{2 \mathrm{h}}$ & $C_{2\mathrm{v}}$ & Assignment \\\hline
        (1$\sigma_\mathrm{u}$)$^2$ & (1b$_\mathrm{1u}$)$^2$ & (1b$_2$)$^2$ & S 1s \\
        (1$\sigma_\mathrm{g}$)$^2$ & (1a$_\mathrm{g}$)$^2$ & (1a$_1$)$^2$ & S 1s \\
        (2$\sigma_\mathrm{g}$)$^2$ & (2a$_\mathrm{g}$)$^2$ & (2a$_1$)$^2$ & C 1s \\
        (2$\sigma_\mathrm{u}$)$^2$ & (2b$_\mathrm{1u}$)$^2$ & (2b$_2$)$^2$ & S 2s \\
        (3$\sigma_\mathrm{g}$)$^2$ & (3a$_\mathrm{g}$)$^2$ & (3a$_1$)$^2$ & S 2s \\
        (3$\sigma_\mathrm{u}$)$^2$ & (3b$_\mathrm{1u}$)$^2$ & (3b$_2$)$^2$ & S 2p \\
        (4$\sigma_\mathrm{g}$)$^2$ & (4a$_\mathrm{g}$)$^2$ & (4a$_1$)$^2$ & S 2p \\
        (1$\pi_\mathrm{g}$)$^4$ & (1b$_\mathrm{2g}$)$^2$, (1b$_\mathrm{3g}$)$^2$ & (4b$_2$)$^2$, (1a$_2$)$^2$ & S 2p \\
        (1$\pi_\mathrm{u}$)$^4$ & (1b$_\mathrm{2u}$)$^2$, (1b$_\mathrm{3u}$)$^2$ & (1b$_1$)$^2$, (5a$_1$)$^2$ & S 2p \\
        (5$\sigma_\mathrm{g}$)$^2$ & (5a$_\mathrm{g}$)$^2$ & (6a$_1$)$^2$ & $\sigma$ \\
        (4$\sigma_\mathrm{u}$)$^2$ & (4b$_\mathrm{1u}$)$^2$ & (5b$_2$)$^2$ & $\sigma$ \\
        (6$\sigma_\mathrm{g}$)$^2$ & (6a$_\mathrm{g}$)$^2$ & (7a$_1$)$^2$ & $\sigma$ \\
        (5$\sigma_\mathrm{u}$)$^2$ & (5b$_\mathrm{1u}$)$^2$ & (6b$_2$)$^2$ & $\sigma$ \\
        (2$\pi_\mathrm{u}$)$^4$ & (2b$_\mathrm{2u}$)$^2$, (2b$_\mathrm{3u}$)$^2$ & (2b$_1$)$^2$, (8a$_1$)$^2$ & $\pi$ \\
        (2$\pi_\mathrm{g}$)$^4$ & (2b$_\mathrm{2g}$)$^2$, (2b$_\mathrm{3g}$)$^2$ & (7b$_2$)$^2$, (2a$_2$)$^2$ & $n$ \\
        \hline
        3$\pi_\mathrm{u}$ & 5b$_\mathrm{2u}$, 3b$_\mathrm{3u}$ & 3b$_1$, 9a$_1$ & $\pi^*$ \\
        7$\sigma_\mathrm{g}$ & 7a$_\mathrm{g}$ & 10a$_1$ & $\sigma^*$ \\
        6$\sigma_\mathrm{u}$ & 6b$_\mathrm{1u}$ & 8b$_2$ & $\sigma^*$ \\
        \hline\hline
    \end{tabular}
\captionsetup{width=15cm}
\caption{Electronic configuration (Hartree-Fock) of the GS in linear and bent geometry.}
\label{tab:GS-HFconfig}
\end{table}

\begin{figure}[htp!]
    \centering
    \includegraphics[scale=0.35]{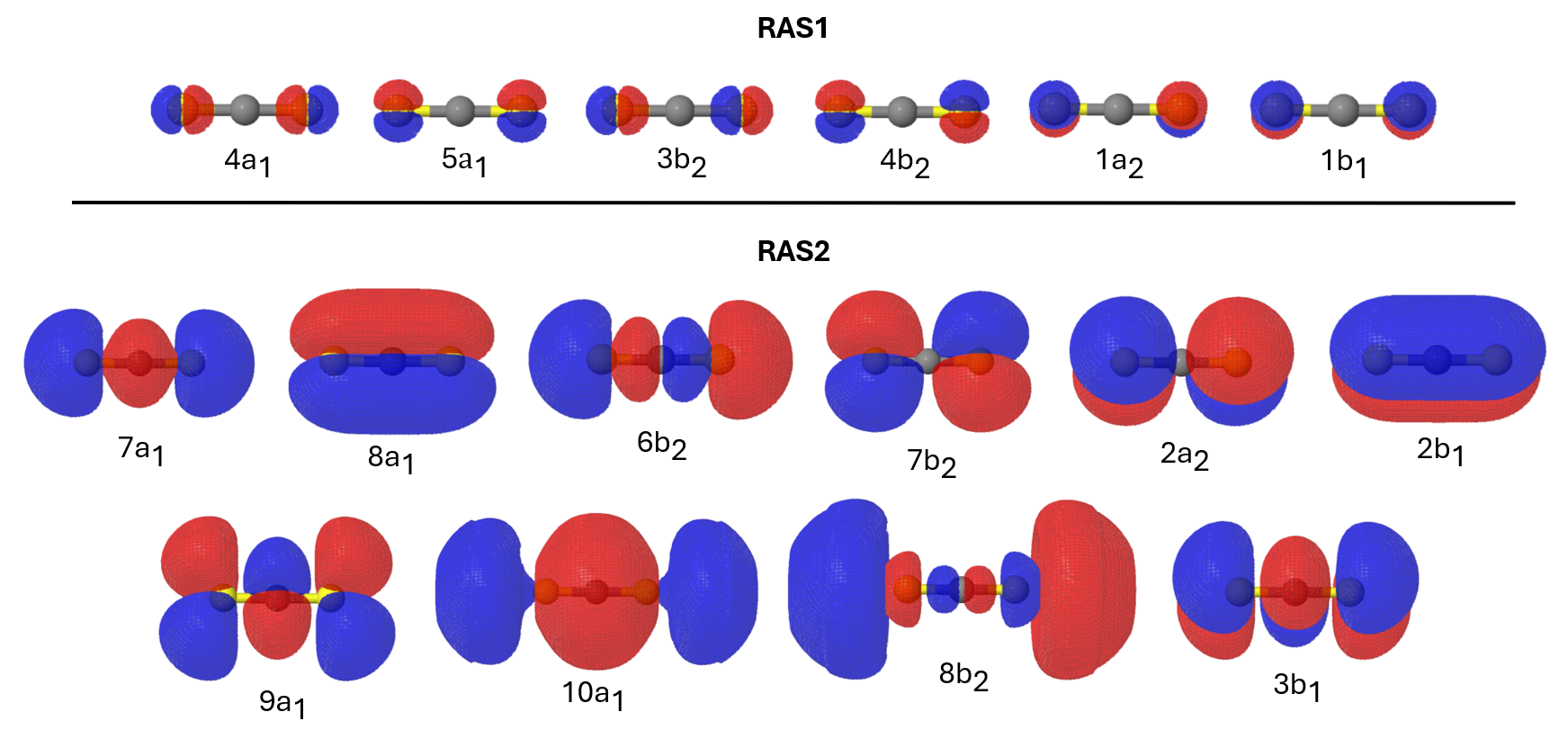}
    \caption{CS$_2$. Active space orbitals of the electronic ground state. The orbitals are SCF orbitals.}
\label{fig:AS_CS2_GS_SCF}
\end{figure}
\begin{figure}[htp!]
    \centering
    \includegraphics[scale=0.5]{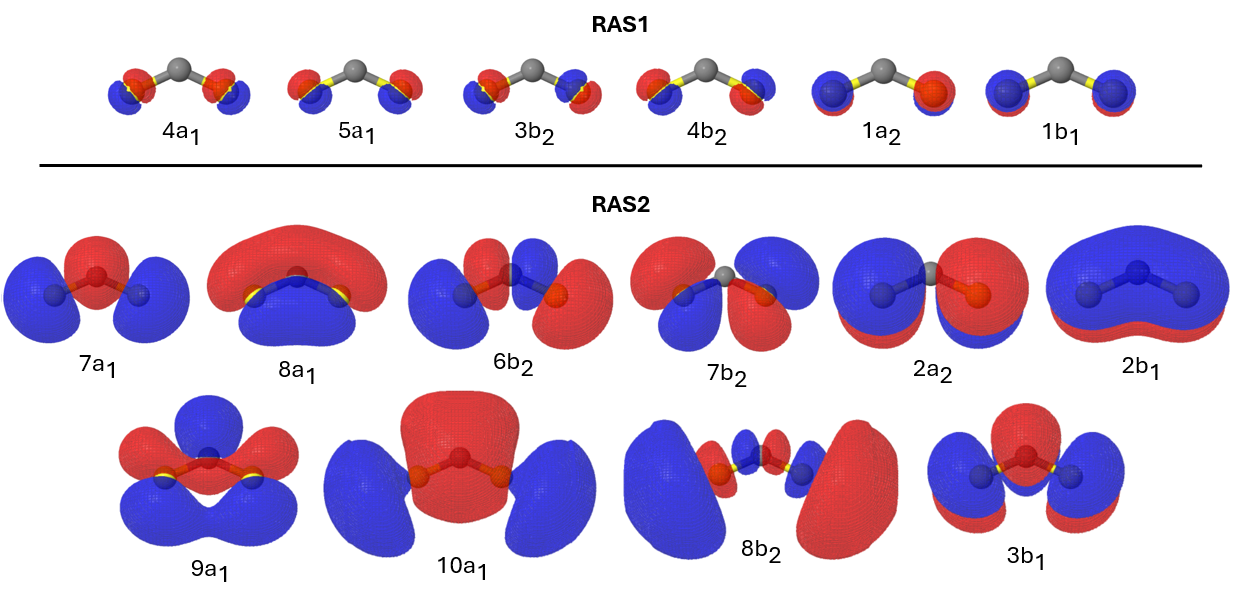}
    \caption{CS$_2$. Active space of 
    $1~{}^{1}\!B_2$ (at the optimized geometry of the state). The orbitals are SCF orbitals.}
    \label{fig:AS_CS2_1_1B2_SCF}
\end{figure}
\begin{figure}[htp!]
    \centering
    \includegraphics[scale=0.5]{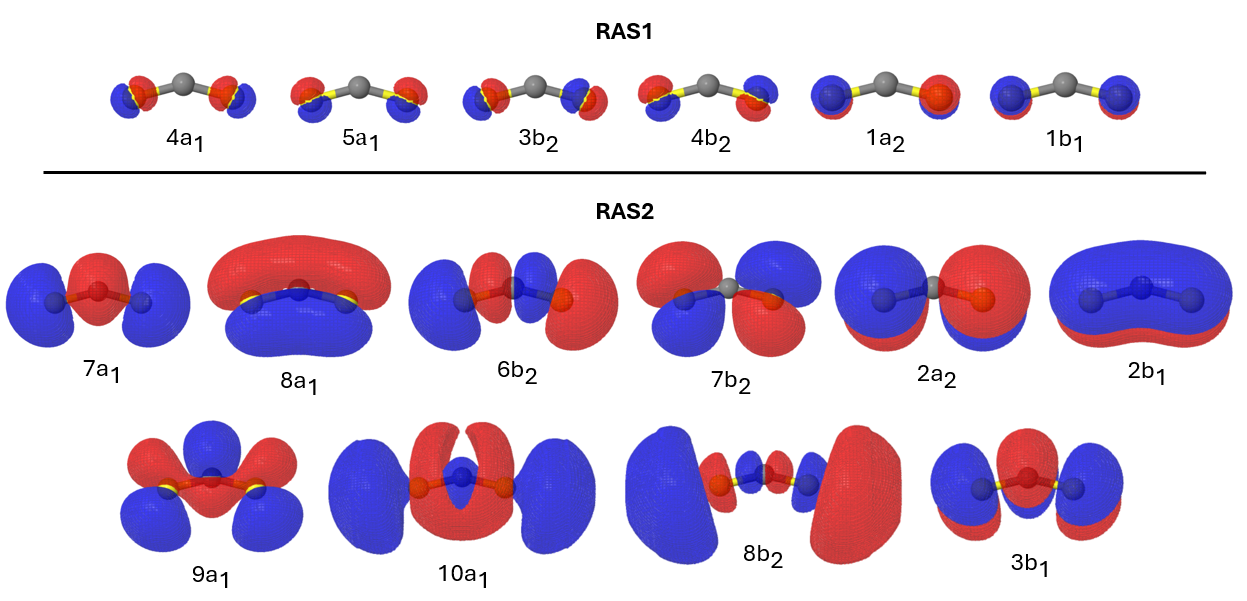}
    \caption{CS$_2$. Active space of $2~{}^{1}\!B_2$ (at the optimized geometry of the state). The orbitals are SCF orbitals.}  \label{fig:AS_CS2_2_1B2_SCF}
\end{figure}
\begin{figure}[htp!]
    \centering
    \includegraphics[scale=0.5]{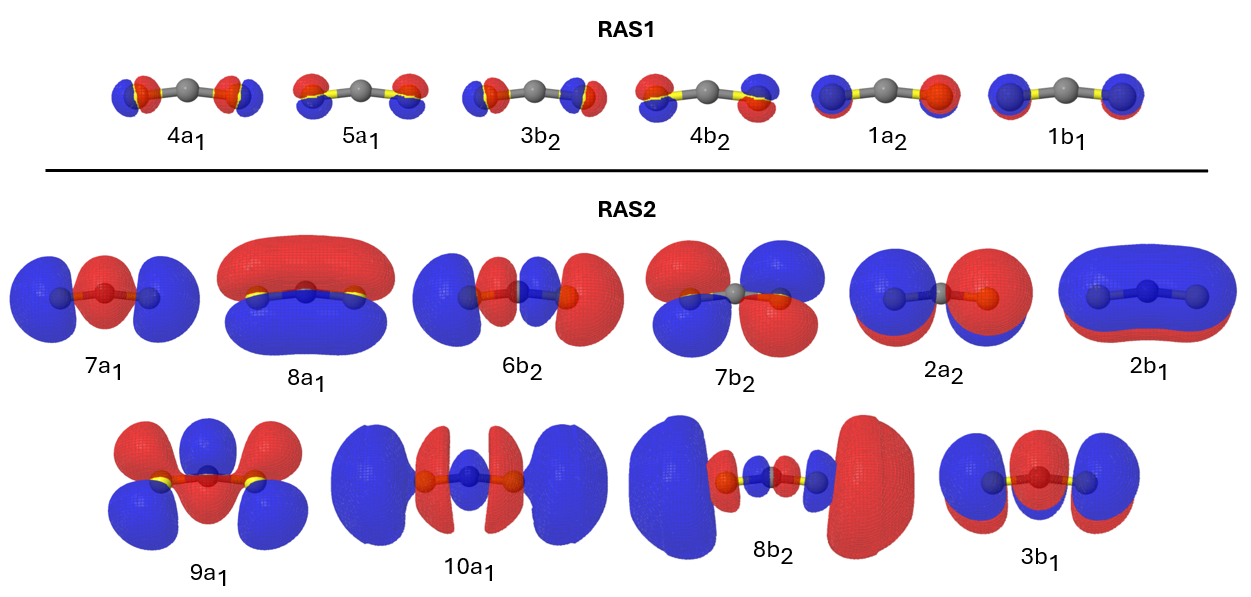}
    \caption{CS$_2$. Active space of $2~{}^{3}\!A_2$ (at the optimized geometry of the state). The orbitals are SCF orbitals.}
\label{fig:AS_CS2_2_3A2_SCF}
\end{figure}
\begin{figure}[htp!]
    \centering
    \includegraphics[scale=0.5]{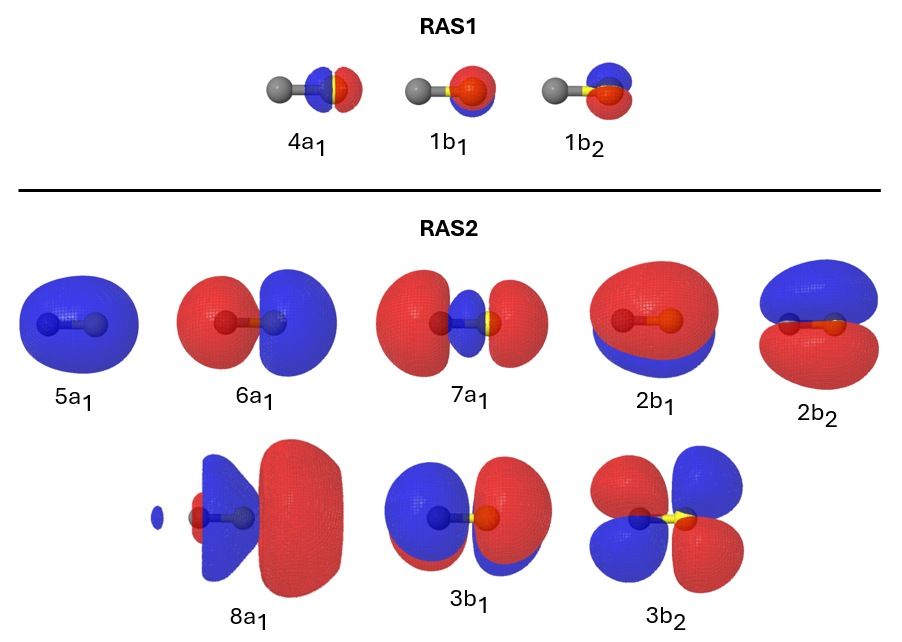}
    \caption{CS. Active space of the electronic ground state. The orbitals are SCF orbitals.}
\label{fig:AS_CS_GS_SCF}
\end{figure}
\begin{figure}[htp!]
    \centering
    \includegraphics[scale=0.29]{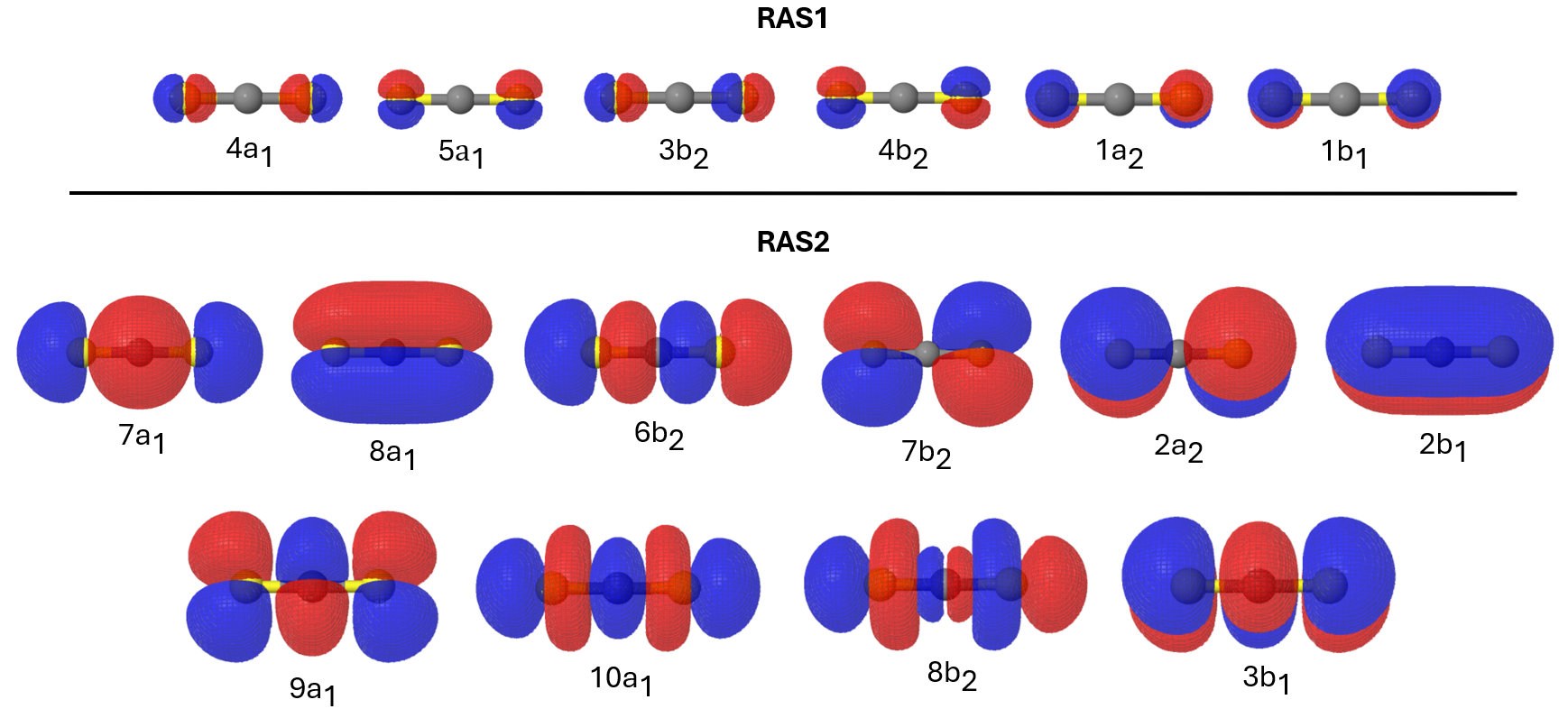}
    \caption{CS$_2$. Active space of the electronic ground state. RASSCF orbitals with SCF labels.}
    \label{fig:AS_CS2_GS_RASSCF}
\end{figure}
\begin{figure}[htp!]
    \centering
\includegraphics[scale=0.5]{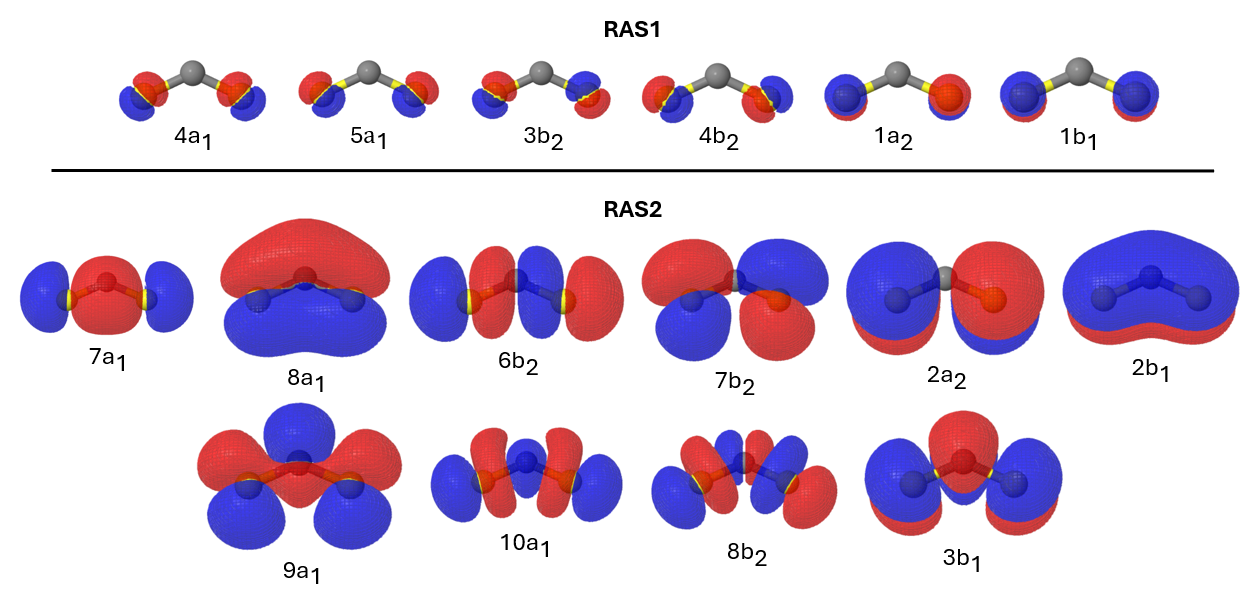}
    \caption{CS$_2$. Active space of $1~{}^{1}B_2$ (at the optimized geometry of the state). RASSCF orbitals with SCF labels.}
\label{fig:AS_CS2_1_1B2_RASSCF}
\end{figure}
\begin{figure}[htp!]
    \centering
\includegraphics[scale=0.5]{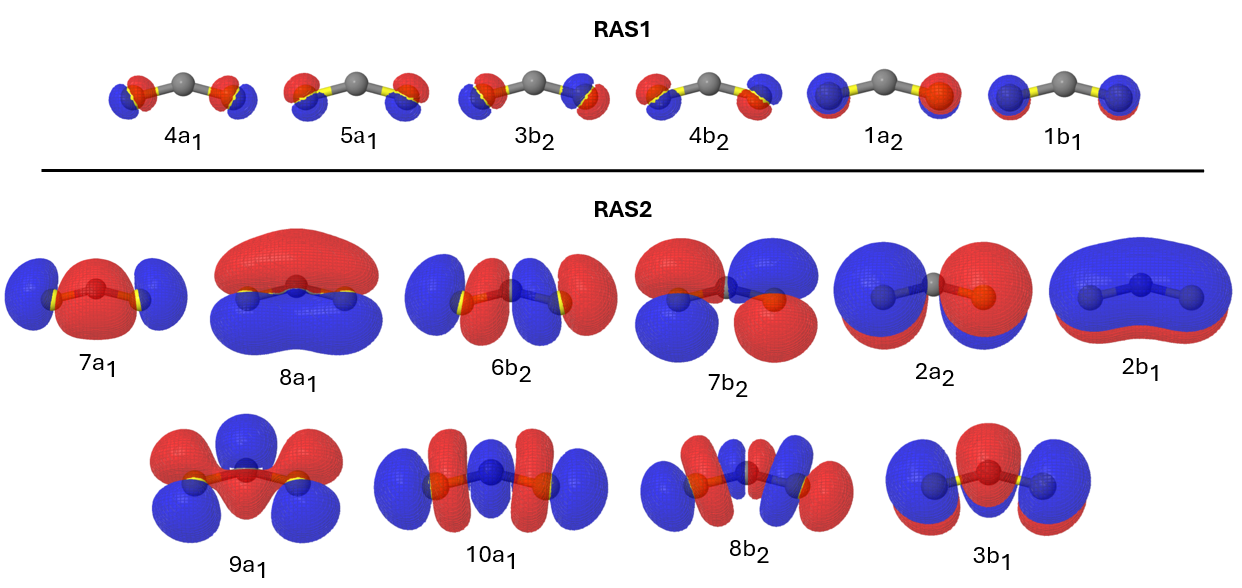}
    \caption{CS$_2$. Active space of $2~{}^{1}B_2$ (at the optimized geometry of the state). RASSCF orbitals with SCF labels.}
\label{fig:AS_CS2_2_1B2_RASSCF}
\end{figure}
\begin{figure}[htp!]
\centering  \includegraphics[scale=0.5]{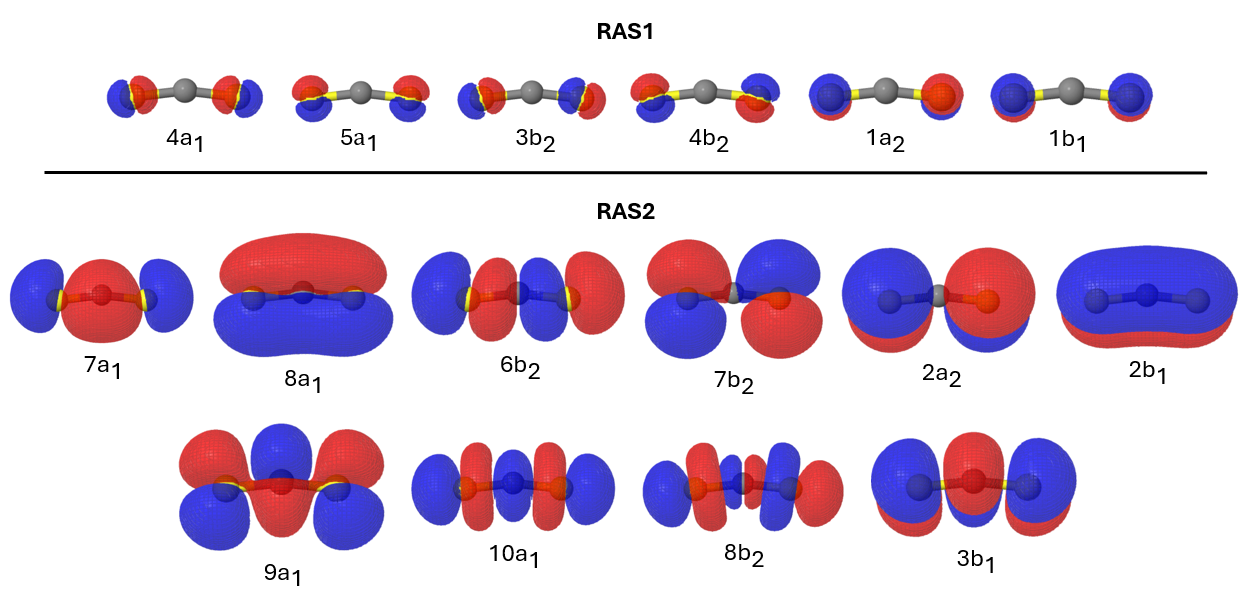}
\caption{CS$_2$. Active space of $2~{}^{3}A_2$ (at the optimized geometry of the state). RASSCF orbitals with SCF labels.}
\label{fig:AS_CS2_2_3A2_RASSCF}
\end{figure}
\begin{figure}[htp!]
    \centering
\includegraphics[scale=0.5]{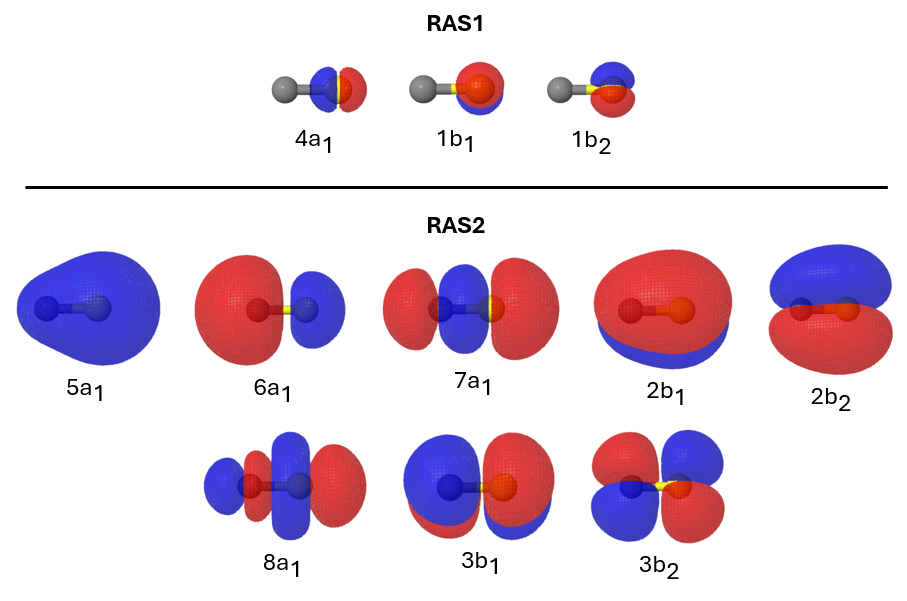}
    \caption{CS. Active space of the electronic ground state. RASSCF orbitals with SCF labels.}
\label{fig:AS_CS_GS_RASSCF}
\end{figure}

We show in Figs.~\ref{fig:AS_CS2_GS_SCF} to \ref{fig:AS_CS2_2_3A2_RASSCF}
the active spaces used in our calculations for the ground state and the valence excited states of CS$_2$. 
The active orbitals of CS are shown in Figs.~\ref{fig:AS_CS_GS_SCF} and \ref{fig:AS_CS_GS_RASSCF}.
The orbitals are plotted with Jmol \cite{jmol} using an MO cutoff and translucent thresholds of 0.02 and 0.25, respectively. We provide both the SCF and RASSCF orbitals since the appearance of the $10\mathrm{a}_{1}$ and $8\mathrm{b}_{2}$ orbitals changes significantly between the two sets of orbitals. We have kept the SCF orbital labels for the RASSCF orbitals.
The basis set used is ANO-RCC-VTZP basis set.
The RAS1 space consists of the six S 2p occupied orbitals (12 electrons). The RAS2 space consisted of 10 orbitals (6 occupied and 4 virtual at SCF level) and 12 electrons. RAS3 is empty.
We allowed for a maximum of one hole in the RAS1 subspace in all cases.

\subsubsection{SCF Orbitals}

\clearpage
\subsubsection{RASSCF Orbitals}

\clearpage

\subsection{Computed XPS spectra}\label{SI3.3}

\begin{figure}[htp!]
    \centering
    \includegraphics[scale=0.55]{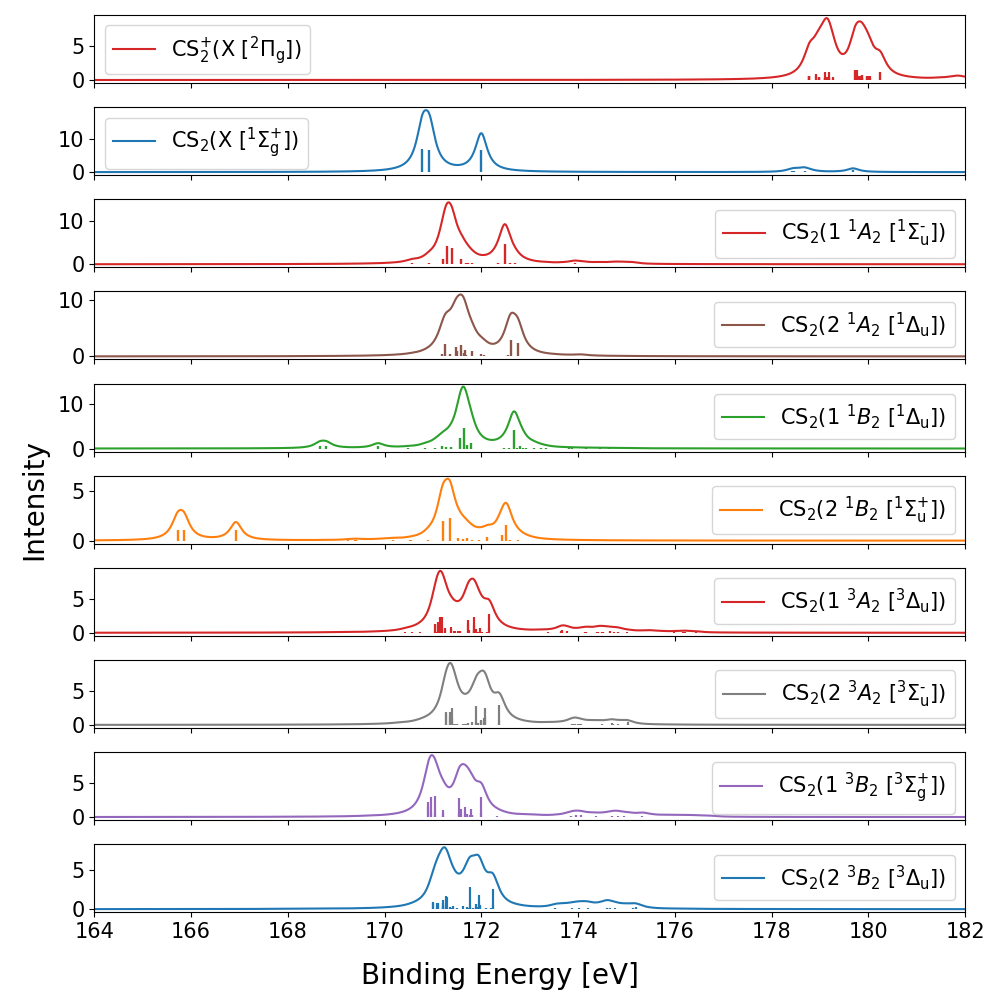}
    \caption{Computed XPS spectra of CS$_2$ and CS$_2^+$. The sticks have been broadened by Lorentzian functions with FWHM = 0.3 eV.}
    \label{fig:cs2_xps}
\end{figure}

\begin{figure}[htp!]
    \centering
\includegraphics[scale=0.55]{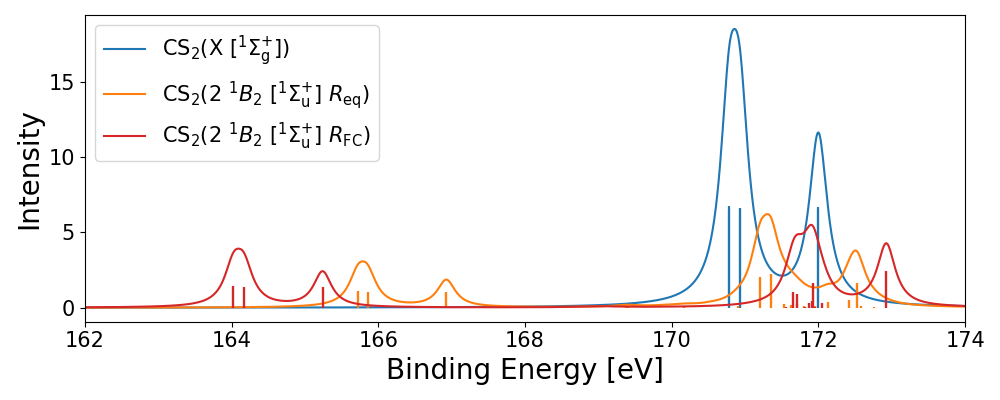}
    \caption{CS$_2$. Transient XPS spectrum of valence-excited ($2~^{1}\!B_{2}~[^{1}\Sigma^{+}_{\mathrm{u}}]$) at both the Franck-Condon (FC) and the optimized (relaxed) excited-state structures. The sticks have been broadened by Lorentzian functions with FWHM = 0.3 eV.}
\label{fig:cs2_fc_v_opt}
\end{figure}

We provide all the theoretical XPS spectra in Fig.~\ref{fig:cs2_xps}. Each spectrum is computed using the optimized structure of the corresponding initial state. The transient XPS spectra of the valence-excited $2~^{1}\!B_{2}~[^{1}\Sigma^{+}_{\mathrm{u}}]$ state computed at both the Franck-Condon (FC) and the optimized (relaxed) excited-state structure are compared in Fig.~\ref{fig:cs2_fc_v_opt}.

\clearpage
\subsection{Assignment of spectroscopic features}
\label{SI3.4}

In Tab.~\ref{tab:valence_transitions_cs2}, we report the valence excitation energies and corresponding oscillator strengths computed at the MS-RASPT2 level of theory, together with the dominant configuration of the transition and the character assigned to the transition. The computed shake-down features are assigned in Tab.~\ref{tab:cs2_shakedown_features}.   

\begin{table}[htp]
\centering

\normalsize
    \begin{tabular}{c|c|c|c|c}
        \hline
        \hline
        Final State & $\Delta$E (eV) & $f_{\mathrm{osc}}$ & Configuration & Character \\
        \hline
        $2 ~{}^{3}A_{2}$ $[{}^{3}\Sigma^{-}_{\mathrm{u}}]$ & 4.01 & - & (2a$_{2}$)$^{-1}$(9a$_{1}$)$^{1}$ [0.44] + (7b$_{2}$)$^{-1}$(3b$_{1}$)$^{1}$ [0.44] & $ \pi^* \leftarrow n $ \\
        $1 ~{}^{1}B_{2}$ $[{}^{1}\Delta_{\mathrm{u}}]$ & 4.10 & - & (7b$_{2}$)$^{-1}$(9a$_{1}$)$^{1}$ [0.43] + (2a$_{2}$)$^{-1}$(3b$_{1}$)$^{1}$ [0.43] & $ \pi^* \leftarrow n $ \\
        $2 ~{}^{1}B_{2}$ $[{}^{1}\Sigma^{+}_{\mathrm{u}}]$ & 6.79 & 1.25 & (7b$_{2}$)$^{-1}$(9a$_{1}$)$^{1}$ [0.42] + (2a$_{2}$)$^{-1}$(3b$_{1}$)$^{1}$ [0.42] & $\pi^* \leftarrow n $ \\
        \hline
        \hline
    \end{tabular}
    \caption{CS$_2$. Computed MS-RASPT2 valence transitions from the ground state.}
    \label{tab:valence_transitions_cs2}
\end{table}

\begin{table}[htp!]
    \centering

\captionsetup{width=10cm}
\normalsize
    \begin{tabular}{c c c}
        \hline
        \hline
        BE (eV) & \textcolor{black}{$R_{\mathrm{FI}}$} & Assignment \\
        \hline
        \vspace{-0.25cm} \\
        \multicolumn{3}{c}{initial state: $\tilde{X}~^{1}\!A_{1} ~[^{1}\Sigma^{+}_{\mathrm{g}}]~R_{\mathrm{FC}}$} \\
        \vspace{-0.25cm} \\
         170.784 & \textcolor{black}{0.664} & 2p$^{-1}$ \\ 
         170.928 & \textcolor{black}{0.660} & 2p$^{-1}$ \\ 
         171.999 & \textcolor{black}{0.662} & 2p$^{-1}$ \\ 
         \hline
         \vspace{-0.25cm} \\
        \multicolumn{3}{c}{initial state: $2 ~^{1}\!B_{2} ~[^{1}\Sigma^{+}_{\mathrm{u}}]~R_{\mathrm{FC}}$} \\
        \vspace{-0.25cm} \\
         164.027 & \textcolor{black}{0.142} & 2p$^{-1}$ \\ 
         164.170 & \textcolor{black}{0.138} & 2p$^{-1}$ \\ 
         165.242 & \textcolor{black}{0.138} & 2p$^{-1}$ \\ 
         \hline
        \vspace{-0.25cm} \\
        \multicolumn{3}{c}{initial state: $2 ~^{1}\!B_{2} ~[^{1}\Sigma^{+}_{\mathrm{u}}]~R_{\mathrm{eq}}$} \\
        \vspace{-0.25cm} \\
         165.723 & \textcolor{black}{0.108} & 2p$^{-1}$ \\ 
         165.856 & \textcolor{black}{0.106} & 2p$^{-1}$ \\ 
         166.928 & \textcolor{black}{0.106} & 2p$^{-1}$ \\ 
         \hline
        \vspace{-0.25cm} \\
        \multicolumn{3}{c}{initial state: $1 ~^{1}\!B_{2} ~[^{1}\Delta_{\mathrm{u}}]~R_{\mathrm{eq}}$} \\
        \vspace{-0.25cm} \\
         168.665 & \textcolor{black}{0.060} & 2p$^{-1}$ \\ 
         168.800 & \textcolor{black}{0.059} & 2p$^{-1}$ \\ 
         169.866 & \textcolor{black}{0.058} & 2p$^{-1}$ \\ 
        \hline
        \hline
    \end{tabular}
    \captionsetup{width=15cm}
    \caption{CS$_{2}$. Computed MS-RASPT2 \textcolor{black}{ground-state primary and valence excited-state} shake-down features (binding energies (BE), \textcolor{black}{averaged Dyson intensities ($R_{\mathrm{FI}}$)}, and corresponding assignment). \textcolor{black}{No shake-down features were found starting from the triplet valence-excited states with a threshold of $10^{-5}$ for the Dyson intensity.}}
    \label{tab:cs2_shakedown_features}
\end{table}

\newpage
\subsection{Additional CCSD results}
\label{SI3.5}

For comparison, we also performed coupled cluster singles and doubles (CCSD) calculations using the aug-cc-pVDZ basis set. All CCSD calculations were carried out with Q-Chem~\cite{Qchem541}.

Table~\ref{tab:linearCS2} collects the CCSD results for the valence excitation energies and oscillator strengths at the linear ground state geometry.
The table also summarizes how the state labels change when changing the conventions for the molecular orientation and point group symmetry.

\begin{figure}[hbt!]
    \centering \includegraphics[width=0.6\linewidth]{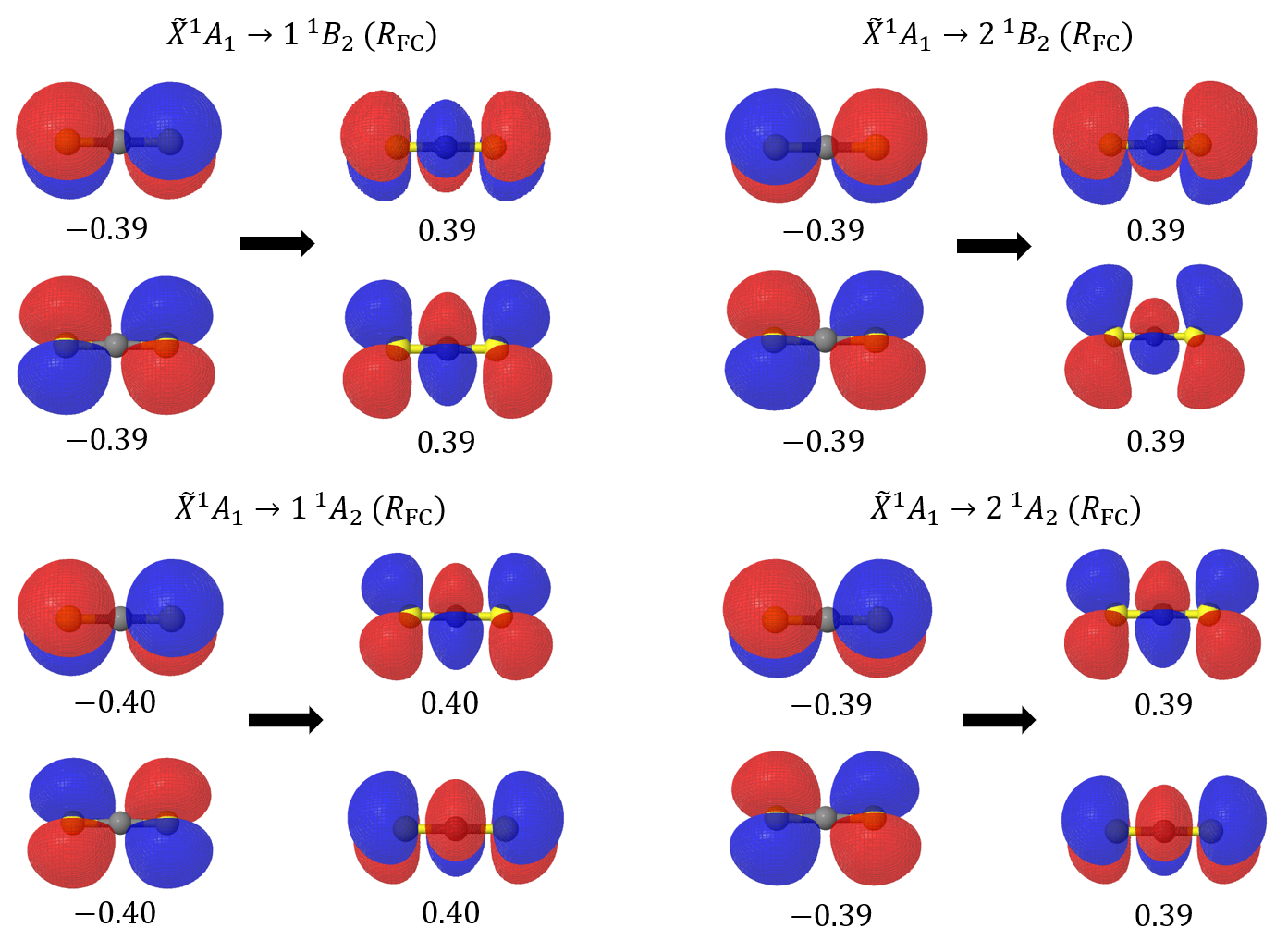}
    \caption{CCSD Natural transition orbitals of the first 4 valence excitations at the linear FC geometry (labelled $R_{\textrm{FC}})$. The  numerical values are the weights of the shown orbital transition in the given NTO. Negative values are for hole orbitals and positive values for particle orbitals.}
    \label{fig:NTO:CCSD:FC}
\end{figure}

\begin{figure}[hbt!]
    \centering \includegraphics[width=0.7\linewidth]{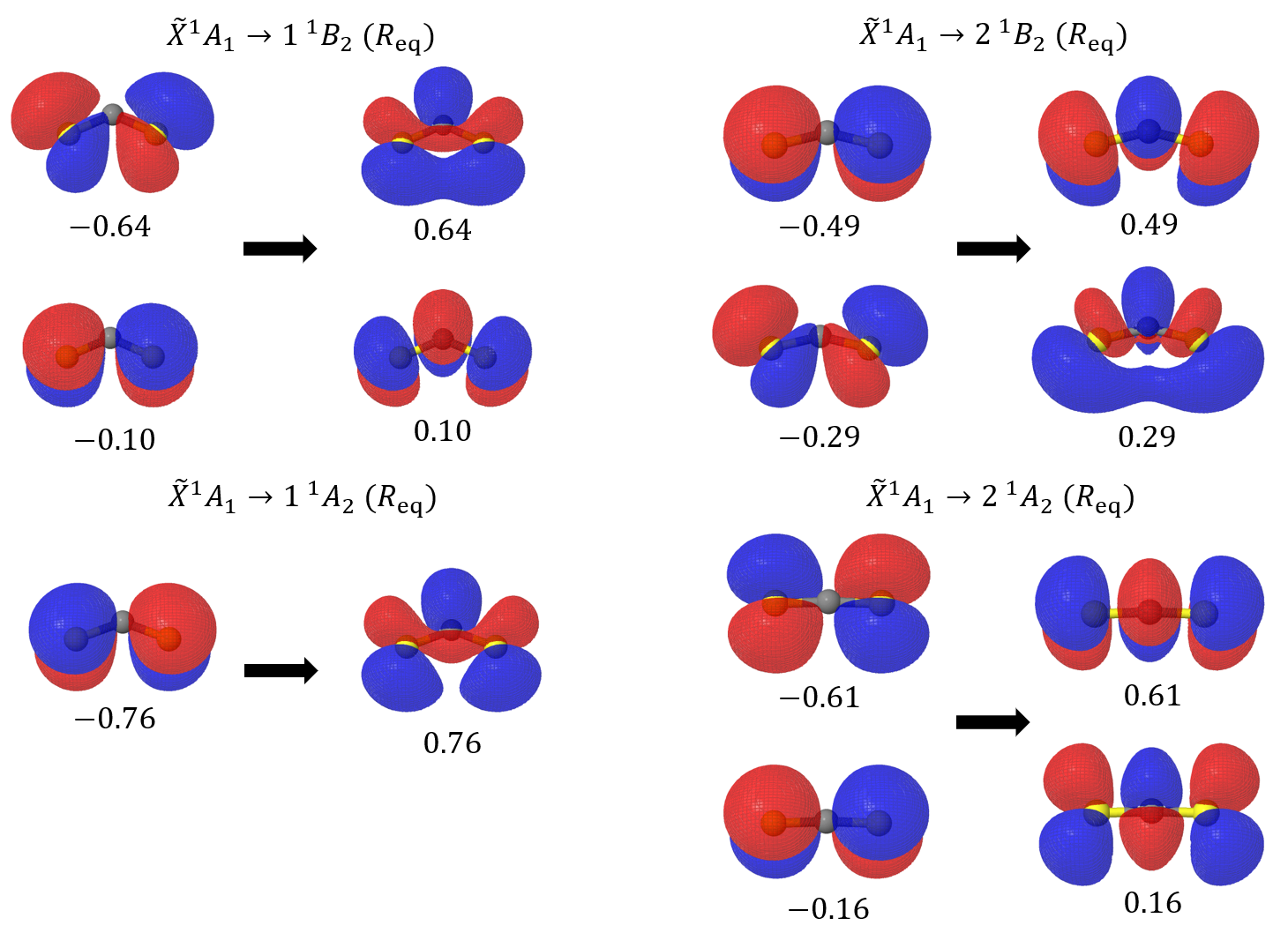}
    \caption{CCSD Natural transition orbitals of the first 4 valence excitations at their respective optimized geometry (labelled $R_{\textrm{eq}}$). The numerical values are the weights of the shown orbital transition in the given NTO. Negative values are for hole orbitals and positive values for particle orbitals.}
    \label{fig:NTO:CCSD:min}
\end{figure}

Figure~\ref{fig:NTO:CCSD:FC} shows the natural transition orbitals (NTO) that characterize the first four singlet valence excited states of CS$_2$ at the linear geometry of the ground state (FC geometry). The NTOs of each of the first four singlet valence excited states at the  respective valence excited state geometry are shown in Fig.~\ref{fig:NTO:CCSD:min}.

Finally, Table~\ref{tab:bentCS2} collects the CCSD/aug-cc-pVDZ excitation energies (and oscillator strengths for the singlet excitations) of the first four singlet and first five triplet excited states at FC geometry and their respective optimized minimum geometries.

\begin{table}[htp]
    \centering

 \captionsetup{width=15cm}
    
    \normalsize
    \begin{tabular}
    {lcl|l|l}
    \hline
{$D_{2h} (z) $} [$D_{\infty h}$] & $\Delta$E~(eV) & $f_{\textrm{osc}}$ (dip)
& $D_{2h}$(y) 
& $C_{2v}$(yz) \\  
\hline
$^1\!A_{\textrm{u}}$ [$^1\Sigma_{\textrm{u}}^-$]      & 4.12081  & 0.0000     & $^1\!A_{\textrm{u}}$ & 
$^1\!A_2$\\
$^1\!B_{1{\textrm{u}}}$ 
[${}^1\Delta_{\textrm{u}}$] 
& 4.15600  & 0.0000   & $^1\!B_{2{\textrm{u}}}$      & $^1\!B_2$ \\
$^1\!A_{\textrm{u}}$ 
[${}^1\Delta_{\textrm{u}}$] & 4.15600  & 0.0000    
& $^1\!A_{\textrm{u}}$        
& $^1\!A_2$ \\
$^1\!B_{1\textrm{u}}$ [$^1\Sigma_{\textrm{u}}^+$] 
& 6.53984  & 1.1739(z) 
& $^1\!B_{2{\textrm{u}}}$(y) 
& $^1\!B_2$(y)\\
$^1\!B_{2{\textrm{g}}}$ [$^1\Pi_{\textrm{g}}$] & 
6.92864  & 0.0000 
& $^1\!B_{1{\textrm{g}}}$ 
& $^1\!B_2$\\
$^1\!B_{3{\textrm{g}}}$ [$^1\Pi_{\textrm{g}}$]
& 6.92864  & 0.0000 
& ${}^1\!B_{3{\textrm{g}}}$ & ${}^1\!A_2$\\
$^1\!B_{1{\textrm{g}}}$ [$^1\Sigma_{\textrm{g}}^-$] 
& 7.41577  & 0.0000 
& $^1\!B_{2{\textrm{g}}}$ 
& $^1\!B_1$\\
$^1\!B_{1{\textrm{g}}}$ [$^1\Delta_{\textrm{g}}$] 
& 7.51654  & 0.0000    
& $^1\!B_{2{\textrm{g}}}$ & $^1\!B_1$\\
$^1\!A_{\textrm{g}}$ [$^1\Delta_{\textrm{g}}$]  & 7.51654  & 0.0000 
& $^1\!A_{\textrm{g}}$ 
& $^1\!A_1$\\
$^1\!B_{2{\textrm{g}}}$ [$^1\Pi_{\textrm{g}}$]& 
7.57573  & 0.0000 & $^1\!B_{1{\textrm{g}}}$     & $^1\!B_2$\\
$^1\!B_{3{\textrm{g}}}$ [$^1\Pi_{\textrm{g}}$] & 7.57573  & 0.0000     & $^1\!B_{3{\textrm{g}}}$     & $^1\!A_2$\\
$^1\!B_{3{\textrm{u}}}$ [$^1\Pi_{\textrm{u}}$] & 
7.90498  & 0.0664(x) & $^1\!B_{3\textrm{u}}$(x) & $^1\!A_1$(z)\\
$^1\!B_{2{\textrm{u}}}$ [$^1\Pi_{\textrm{u}}$]& 7.90498  & 0.0664(y) 
& $^1\!B_{1{\textrm{u}}}$(z) 
& $^1\!B_1$(x)\\
\hline
    \end{tabular}
    \caption{CCSD/aug-cc-pVDZ excitation energies ($\Delta$E) and oscillator strengths ($f_{\textrm{osc}}$) for the linear ground state geometry. Symmetry relabeling according to different orientations of the molecule in $D_{2h}$ and according to $C_{2v}$ (Mulliken's convention). In $D_{2h}$(z)~[$D_{\infty h}]$ the molecule is aligned along z. 
In  $D_{2h}$(y) it is aligned along $y$. In $C_{2v}$ it is aligned along $y$ (such that the bent $C_{2v}$ molecule lies on the yz plane with the z axis as the C$_2$ axis.}
\label{tab:linearCS2}
\end{table}

\begin{table}[]
    \centering
    \scriptsize

    \begin{tabular}{lcl|lc|lc|lc|lc}
 \\  
\hline
\multicolumn{3}{c|}{@FC} 
& 
\multicolumn{2}{c|}{@min(1~${}^1\!A_2$)} 
&
\multicolumn{2}{c|}{@min(1~${}^1\!B_2$)} 
&
\multicolumn{2}{c|}{@min(2~${}^1\!B_2$)} 
&
\multicolumn{2}{c}{@min(2~${}^1\!A_2$)}
\\
\hline
1~${}^1\!A_2$ & 
4.12081 & 0.0000 &
2.20950 & 0.0000 &
1.82458 & 0.0000 &
2.72980 & 0.0000 &
3.69193 & 0.0000
\\
1~${}^1\!B_2$ & 4.15600 & 0.0000 &
2.38444 & 0.0258 &
1.92929 & 0.0274 &
2.92306 & 0.0133 &
3.71273 & 0.0000\\
2~${}^1\!A_2$ & 
4.15600 & 0.0000 & 
3.14745 & 0.0000 &
2.92882 & 0.0000 &
3.31698 & 0.0000 &
3.71488 & 0.0000 
\\
2~${}^1\!B_2$(y) 
& 6.53984 & 1.1739 
& 4.80220 & 0.4509 
& 4.57333 & 0.3648 
& 5.04621 & 0.5782 
& 6.04899 & 1.1253 \\
\hline
& @FC  & & \multicolumn{2}{c|}{@min(1~$^3\!A_2$)} & \multicolumn{2}{c|}{@min(1~$^3\!B_1$)}
& 
\multicolumn{2}{c|}{@min(2~$^3\!B_2$)} &
\multicolumn{2}{c}{@min(2~${}^3\!A_2$)}  
\\
\hline
${}^3\!B_2$ 
& 3.42464 & 
& 1.43616 & 
& 0.66059 & 
& 3.00244 & 
& 2.93212 & \\
${}^3\!A_2$ 
& 3.81557 & 
& 1.92986 & 
& 1.35161 & 
& 3.35497 & 
& 3.27941 & 
\\
${}^3\!B_2$ 
& 3.81557 & 
& 2.98703 & 
& \bf{2.8521} & 
& 3.38450 & 
& 3.33690 & 
\\
${}^3\!A_2$ 
& 4.12345 & 
& 3.00346 & 
& 2.66332 & 
& 3.65409 & 
& 3.60005 & 
\\
${}^3\!A_1$ 
& 6.22134 & 
& 4.20605 & 
& 3.85602 & 
& 5.35411 & 
& 5.27012
%
\\\hline
    \end{tabular}
    \caption{CCSD/aug-cc-pVDZ results for the first 4 singlet and the first 5 triplet excited states at different geometries. $C_{2v}$(yz) symmetry. Symmetry labels follow Mulliken's convention.}
    \label{tab:bentCS2}
\end{table}

\clearpage

\bibliography{SIbib}